\documentclass[fleqn,usenatbib,useAMS]{mnras}

\usepackage{mathptmx}

\usepackage[T1]{fontenc}
\usepackage{ae,aecompl}

\usepackage{graphicx}
\usepackage[section]{placeins}
\DeclareGraphicsExtensions{.pdf,.png,.jpg,.eps}
\usepackage{epstopdf}
\usepackage{color}
\usepackage{amsmath}
\usepackage{amssymb}
\usepackage{amsbsy}
\usepackage{comment}
\usepackage[normalem]{ulem}
\usepackage{enumitem}

\newcommand{\green}[1]{\iffalse #1 \fi}
\usepackage{pbox}

\definecolor{blue}{rgb}{0,0,1}

\usepackage{graphicx}	
\usepackage{amsmath}	
\usepackage{amssymb}	
\usepackage{bm}		
\usepackage{pdflscape}	
\usepackage{color}
\usepackage{exscale}
\usepackage{natbib}
\usepackage{rotating}
\usepackage{rotate}
\usepackage{afterpage}
\usepackage{booktabs,threeparttable}
\usepackage{relsize}
\usepackage{lscape}
\usepackage{tabularx}
\usepackage{longtable}
\usepackage{ltxtable}
\usepackage{txfonts}
\usepackage{bbm}
\usepackage{afterpage}
\usepackage{widetext}
\usepackage{bbold}
\usepackage{footnote}

\def\apjl{ApJ}%
\def\apjs{ApJS}%
\def\ao{Appl.~Opt.}%
\def\aap{A\&A}%
\def\mnras{MNRAS}%
\newcommand{\bea}{\begin{eqnarray}}
\newcommand{\be}{\begin{equation}}
\newcommand{\ben}{\begin{enumerate}}
\newcommand{\bi}{\begin{itemize}}
\newcommand{\eea}{\end{eqnarray}}
\newcommand{\ee}{\end{equation}}
\newcommand{\ei}{\end{itemize}}
\newcommand{\een}{\end{enumerate}}

\usepackage{titlesec}
\titlespacing*{\paragraph}
{0pt}{0.2cm}{0.3cm}

\title[precision matrix expansion]{Precision matrix expansion -- efficient use of numerical simulations in estimating errors on cosmological parameters}

\author[Friedrich, Eifler]{Oliver Friedrich$^{1,2}$ \thanks{E-mail: oliverf@usm.uni-muenchen.de}, Tim Eifler$^{3,4}$ \thanks{E-mail: tim.eifler@jpl.nasa.gov} \\
$^{1}$ Universit\"ats-Sternwarte, Fakult\"at f\"ur Physik, Ludwig-Maximilians Universit\"at M\"unchen, Scheinerstr. 1, 81679 M\"unchen, Germany\\
$^{2}$ Max Planck Institute for Extraterrestrial Physics, Giessenbachstrasse, 85748 Garching, Germany\\
$^{3}$ Jet Propulsion Laboratory, California Institute of Technology, Pasadena, CA 91109, USA\\
$^{4}$ Department of Physics, California Institute of Technology, Pasadena, CA 91125, USA}

\begin{document}

\maketitle

\label{firstpage}

\begin{abstract}
Computing the inverse covariance matrix (or precision matrix) of large data vectors is crucial in weak lensing (and multi-probe) analyses of the large scale structure of the universe. Analytically computed covariances are noise-free and hence straightforward to invert, however the model approximations might be insufficient for the statistical precision of future cosmological data. Estimating covariances from numerical simulations improves on these approximations, but the sample covariance estimator is inherently noisy, which introduces uncertainties in the error bars on cosmological parameters and also additional scatter in their best fit values. For future surveys, reducing both effects to an acceptable level requires an unfeasibly large number of simulations.

In this paper we describe a way to expand the true precision matrix around a covariance model and show how to estimate the leading order terms of this expansion from simulations. This is especially powerful if the covariance matrix is the sum of two contributions, $\smash{\mathbf{C} = \mathbf{A}+\mathbf{B}}$, where $\smash{\mathbf{A}}$ is well understood analytically and can be turned off in simulations (e.g. shape-noise for cosmic shear) to yield a direct estimate of $\smash{\mathbf{B}}$. We test our method in mock experiments resembling tomographic weak lensing data vectors from the Dark Energy Survey (DES) and the Large Synoptic Survey Telecope (LSST). For DES we find that $400$ N-body simulations are sufficient to achive negligible statistical uncertainties on parameter constraints. For LSST this is achieved with $2400$ simulations. The standard covariance estimator would require >$10^5$ simulations to reach a similar precision. We extend our analysis to a DES multi-probe case finding a similar performance.
\end{abstract}


\renewcommand{\thefootnote}{\arabic{footnote}}
\setcounter{footnote}{0}

\section{Introduction}
\label{sec:introduction}

\noindent Wide area surveys such as the currently running Dark Energy Survey \citep[DES,][]{Flaugher2005} or the upcoming Large Synoptic Survey Telecope \citep[LSST,][]{LSST2009} will collect vast amounts of data about the large scale structure on the universe. In cosmological analyses this data can e.g. be compressed into measurements of 2-point correlation functions of galaxy clustering or cosmic shear. In a redshift-tomographic analysis this will easily accumulate to data vectors with several hundreds of data points. Testing cosmological models from a measurement of such a large data vector requires precise knowledge of the inverse covariance matrix of the noise in this data vector. There has been extensive research on the impact of errors associated with covariance estimation on the constraints derived on cosmological parameters. \citet{Hartlap2007} discussed the fact that the inverse of an unbiased covariance estimator is not an unbiased estimator for the inverse covariance matrix (\emph{the precision matrix}). They also described a way to correct for this when assuming that the covariance estimate follows a Wishart distribution (see also \citealt{Kaufman} and \citealt{Anderson2003}). The noise properties of this corrected precision matrix estimator and its impact on the constraints derived on cosmological parameters was e.g. investigated by \citet{Taylor2013, Dodelson2013, Taylor2014}.

\citet[hereafter SH16a]{Sellentin2016a} have presented a different approach: given a covariance estimate they marginalize over the posterior distribution of the true precision matrix to compute the likelihood in parameter space. Assuming that the covariance estimate follows a Wishart distribution they have derived a simple, closed-form expression for the resulting likelihood function. In \citet{Sellentin2016b} they have extended these results to derive the information loss in parameter space due to noisy covariance estimates. A fully non-Gaussian treatment of the effects discussed in \citet[hereafter DS13]{Dodelson2013} is however still missing.

Prior knowledge on the sparsity of the covariance matrix and the precision matrix was used by \citet{Paz2015} and \citet{Padmanabhan2015} to improve estimates of the precision matrix from few simulations. \citet{Pope2008} investigated shrinkage estimators of the covariance, i.e. a mixing of estimated and modelled covariance matrices. This however raises the task of finding an equivalent to the Kaufman-Hartlap correction for such a mixture of estimated and analytic matrices. More recently, \citet{Joachimi2017} describes a non-linear extension of that estimator which combines covariance estimates from two sets of independent data vector realisations and hence does not require a covariance model.

In this paper we describe a way to expand the true precision matrix around a covariance model as a power series in the deviation between model and true covariance. Assuming a Wishart realisation for the true covariance and using the results on invariant moments of the Wishart distribution by \citet{Letac2004} we derive an unbiased estimator for the up to second order expansion of the true precision matrix. This becomes especially powerful if parts of the covariance matrix that are well understood analytically can be turned off in simulations in order to yield a direct estimate of the remaining covariance parts. In Sect. \ref{sec:method} we recap the main problems of estimating parameter constraints from noisy covariance estimates and present our method of "Precision Matrix Expansion" (PME). In Sect. \ref{sec:examples} we perform numerical experiments that mimic data from the Dark Energy Survey (DES) and the Large Synoptic Survey Telecope (LSST) likelihood analyses to test the performance of our idea. Sect. \ref{sec:conclusions} concludes with a discussion of our results.


\section{Parameter constraints from noisy covariance estimates}
\label{sec:constraints_from_noisy_cov}

\begin{figure*}
  \includegraphics[width=8.5cm]{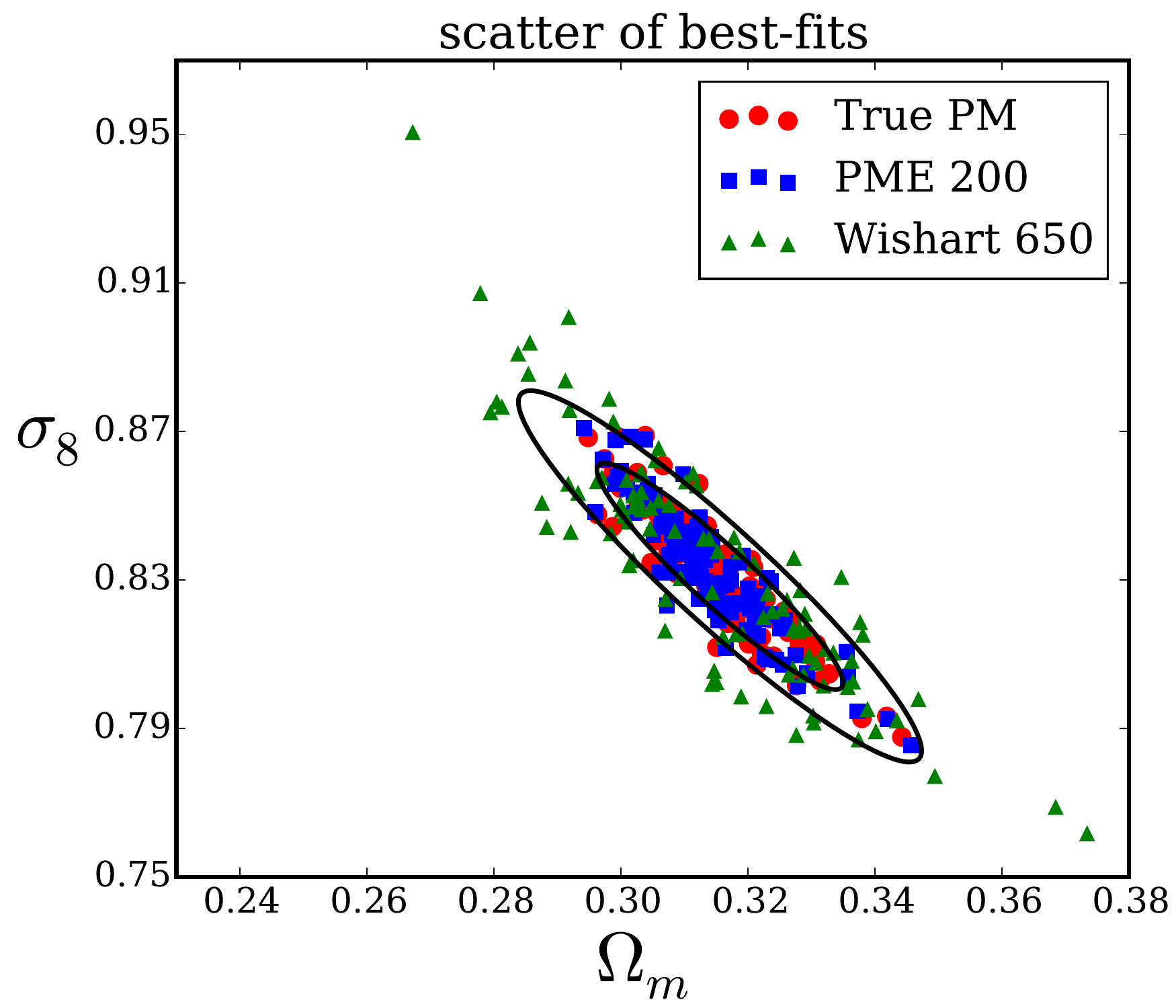}  \includegraphics[width=8.5cm]{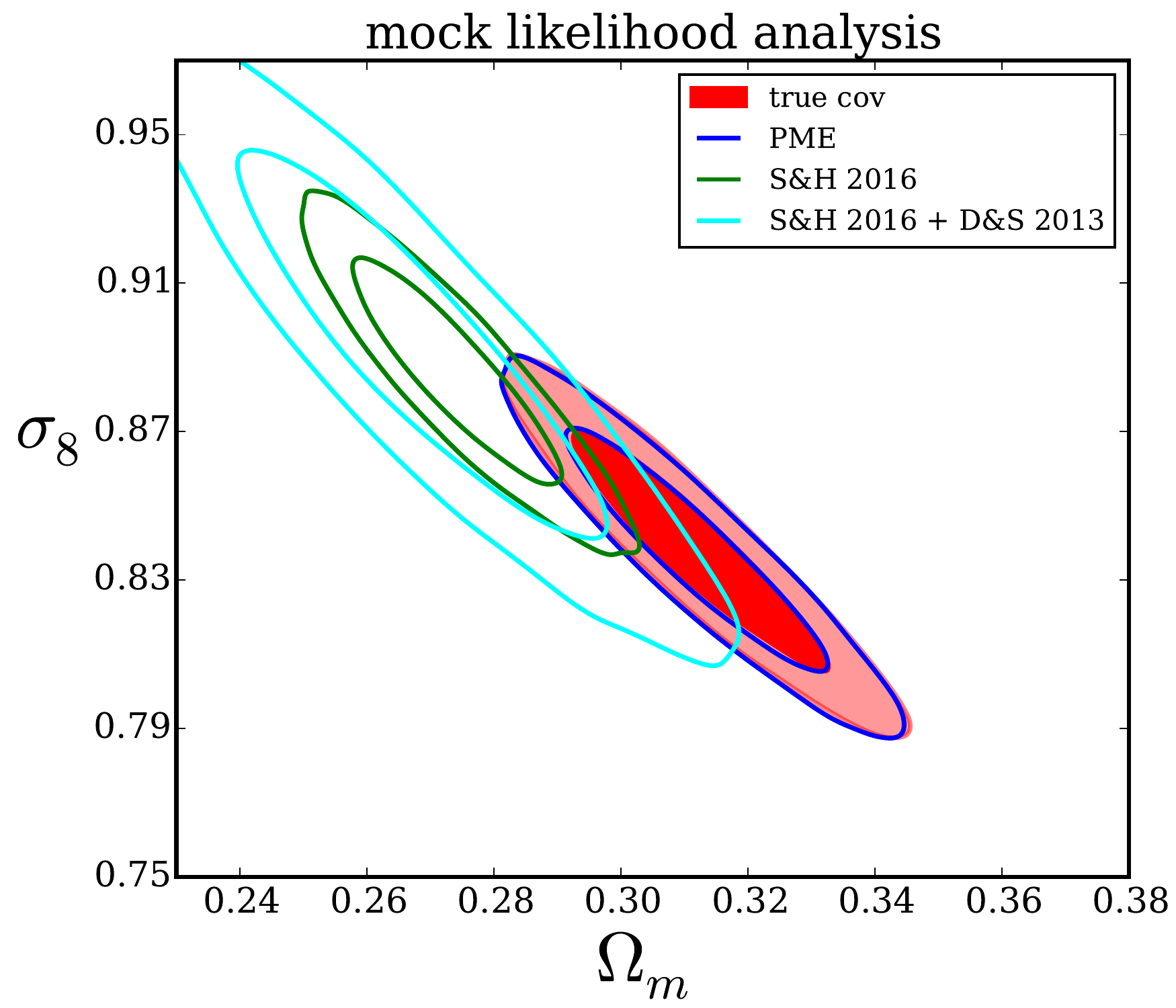}
   \caption{Left: Best fit parameter pairs $(\Omega_m, \sigma_8)$ obtained from random realisations of a DES-like weak lensing data vector with $450$ data points when using different approaches to compute the precision matrix. The red points assume that the true covariance matrix is known while for the green points we draw a Wishart realisation of the covariance ($N_s = 450 + 200 = 650$ simulations) for each data vector. The blue points are obtained with the method of precision matrix expansion (and allowing only $200$ simulations to estimate the expansion). The black contours display the 1$\sigma$ and 2$\sigma$ Fisher contours derived from our fiducial covariance.
   Right: For one of the random realisations we perform a complete likelihood analysis and show the 1$\sigma$ and 2$\sigma$ contours in the $\Omega_m-\sigma_8$ plane after marginalizing over $w_0$ and $w_a$ (see Sec. \ref{sec:examples} for details). The contours obtained from the Wishart realisation of the covariance are clearly offset from those obtained from the true covariance matrix. We recommend to account for this by expanding the likelihood around its maximum (of the full parameter space, which in this figure is 4-dimensional) with the factor derived by DS13. This leads to a decreased contraining power of our mock survey. The use of PME manages to significantly decrease this contour offset.}
  \label{fi:DES_CS_seed1_200}
\end{figure*}

We begin by outlining the main task of this paper. Let $\smash{\boldsymbol{\hat \xi}}$ be a vector of $N_d$ data points measured from observational data and let $\boldsymbol{\xi}[\boldsymbol{\pi}]$ be a model for this data vector that depends on a vector of $N_p$ parameters $\boldsymbol{\pi}$. If $\mathbf{C}$ is the covariance matrix of $\smash{\boldsymbol{\hat \xi}}$ then a standard way to constrain the parameters $\boldsymbol{\pi}$ is to assign a posterior distribution $p(\boldsymbol{\pi}|\boldsymbol{\hat \xi})$ to them as
\begin{equation}
\label{eq:standard_likelihood}
p(\boldsymbol{\pi}|\boldsymbol{\hat \xi}) \sim \exp\left( -\frac{1}{2} \chi^2\left[\boldsymbol{\pi}\ |\ \boldsymbol{\hat \xi}, \mathbf{C} \right] \right)\ p(\boldsymbol{\pi})
\end{equation}
with
\begin{equation}
\label{eq:chi_squared}
\chi^2\left[\boldsymbol{\pi}\ |\ \boldsymbol{\hat \xi}, \mathbf{C} \right] = \left(\boldsymbol{\hat \xi} - \boldsymbol{\xi}[\boldsymbol{\pi}] \right)^T \mathbf{C}^{-1} \left(\boldsymbol{\hat \xi} - \boldsymbol{\xi}[\boldsymbol{\pi}] \right)
\end{equation}
and $p(\boldsymbol{\pi})$ being a prior density incorporating apriori knowledge or assumptions on $\boldsymbol{\pi}$. These expressions in fact ignore that $\mathbf{C}$ also can be depedent on $\boldsymbol{\pi}$. We will do this throughout this paper and refer the reader to \cite{Eifler2009} who investigated the impact of cosmology dependent covariance matrices on cosmic shear likelihood analyses. Another assumption that goes into Eq. \ref{eq:standard_likelihood} is that the measured data vector $\smash{\boldsymbol{\hat \xi}}$ is drawn from a multi-variate Gaussian distribution. In wide area surveys this is justified in the limit where one can consider the survey to consist of many independent sub-regions, such that the measurements in those regions add up to a Gaussian data vector by means of the central limit theorem.

If the covariance matrix $\mathbf{C}$ is not exactly known, it can e.g. be estimated from N-body simulations. If $\smash{\boldsymbol{\hat \xi}_i}$, $i = 1 ... N_s$, are a number of independent measurements of $\boldsymbol{\xi}$ in simulations then an unbiased estimate of $\mathbf{C}$ is given by
\begin{equation}
\label{eq:cov_estimator}
\mathbf{\hat C} := \frac{1}{\nu}\sum_{i=1}^{N_s} \left(\boldsymbol{\hat \xi}_i-\boldsymbol{\bar \xi}\right)\left(\boldsymbol{\hat \xi}_i-\boldsymbol{\bar \xi}\right)^T\ ,
\end{equation}
where $\nu = N_s-1$ and $\boldsymbol{\bar \xi}$ is the sample mean of the $\smash{\boldsymbol{\hat \xi}_i}$. We will assume $\smash{\mathbf{\hat C}}$ to have a Wishart distribution with $\nu$ degrees of freedom which follows from our assumption that $\smash{\boldsymbol{\hat \xi}}$ and the $\smash{\boldsymbol{\hat \xi}_i}$ are Gaussian distributed (cf. \citealt{Taylor2013}).

To compute the likelihood in Eq. \ref{eq:standard_likelihood} we need to know the precision matrix, i.e. is the inverse covariance matrix $\smash{\mathbf{\Psi} = \mathbf{C}^{-1}}$. According to \citeauthor{Kaufman} (1967, see also \citealt{Hartlap2007, Taylor2013}) an unbiased estimator for $\smash{\mathbf{\Psi}}$ can be constructed from $\smash{\mathbf{\hat C}}$ as
\begin{equation}
\mathbf{\hat \Psi} = \frac{\nu - N_d - 1}{\nu} \mathbf{\hat C}^{-1}
\end{equation}
and we will call the factor of $(\nu - N_d - 1)/\nu$ the Kaufman-Hartlap-correction. 

Given a measurement $\smash{\boldsymbol{\hat \xi}}$ of the data vector one can derive the posterior density of the model parameters $p(\boldsymbol{\pi}|\boldsymbol{\hat \xi})$ by means of equations \ref{eq:standard_likelihood} and \ref{eq:chi_squared}. A noisy precision matrix estimate influences this inference in two ways:
\begin{itemize}
\item it adds noise to the width of likelihood contours derived from inserting the precision matrix estimate into the figure of merit $\chi^2$ {(Eq.~\ref{eq:chi_squared})}.
\item it adds noise to the location of likelihood contours. E.g. the maximum likelihood estimator for the parameters $\boldsymbol{\pi}$ would be
\begin{equation}
\label{eq:best_fit}
\boldsymbol{\hat \pi}_{\mathrm{ML}} = \underset{\boldsymbol{\pi}}{\min} \left\lbrace\left(\boldsymbol{\hat \xi} - \boldsymbol{\xi}[\boldsymbol{\pi}] \right)^T \mathbf{\hat \Psi} \left(\boldsymbol{\hat \xi} - \boldsymbol{\xi}[\boldsymbol{\pi}] \right)\right\rbrace\ .
\end{equation}
When using a noisy precision matrix the uncertainties of $\boldsymbol{\hat \pi}_{\mathrm{ML}}$ have contributions from both the noise in $\smash{\boldsymbol{\hat \xi}}$ and the noise in $\smash{\mathbf{\hat \Psi}}$.
\end{itemize}

The astro-statistics literature has so far focused on the first effect, i.e. on the uncertainties on contour width due to noise in the estimate $\smash{\mathbf{\hat\Psi}}$ \citep{Taylor2013, Taylor2014, Sellentin2016a, Sellentin2016b}. \citet{Sellentin2016b} provide the most complete demonstration that $\smash{\mathbf{\hat \Psi}}$ yields a good estimate of the width of the posterior contours as long as $ N_s - N_d \gg N_p$.

The more critical effect however is the additional noise of $\boldsymbol{\hat \pi}_{\mathrm{ML}}$. DS13 (also see appendix \ref{app:Dodelson2013}) showed that the uncertainty on the position of likelihood contours from noise in $\smash{\mathbf{\hat\Psi}}$ is only negligible if $N_s - N_d \gg N_d - N_p$ which is a much more demanding criterion for current cosmological data vectors. We demonstrate this in the left-hand panel of Fig. \ref{fi:DES_CS_seed1_200}, where we show 100 randomly drawn realisations of a DES-like weak lensing data vector with $N_d = 450$ and a halo model covariance matrix (see Sec. \ref{sec:examples} for further details). For each of the 100 data vectors we have also generated Wishart realisations of our covariance matrix corresponding to an estimate from $N_s = 650$ simulations. Using either the true covariance or the estimated one, we then determine the best fitting parameters $\Omega_m$ and $\sigma_8$ (after marginalizing over equation-of-state parameters of dark energy, $w_0$ and $w_a$). The best-fits obtained from a noisy covariance (green points) clearly display a much larger scatter than those obtained from the true covariance (red points). Also shown are the best fits obtained by precision matrix expansion (PME, blue points) which we are going to introduce in the next section. Here we assumed that only $N_s = 200$ simulations are available to estimate the PME, which gives best fit values that are significantly closer to the ones obtained when knowing the true covariance matrix.

When reconstructing $p(\boldsymbol{\pi}|\boldsymbol{\hat \xi})$ (e.g. from a Monte-Carlo-Markov-Chain) this can lead to significant offsets between likelihood contours inferred from the true covariance matrix and likelihood contours inferred from a covariance estimate -- even if the overall width of the likelihood contours is captured well by the covariance estimate. We demonstrate this in the right-hand panel of figure \ref{fi:DES_CS_seed1_200}. DS13 have derived a factor (see appendix \ref{app:Dodelson2013}) by which parameter contours obtained from a Wishart realisation of the covariance should be expanded in order to account for this additional scatter. However, their derivation relies on the assumption of a Gaussian parameter likelihood and is only applicable to the extent that a Fisher analysis is accurate. The current state of the art for dealing with noisy covariance estimates is hence a combination of SH16a and DS13: expanding the contours derived from the SH16a likelihood by the DS13 factor. We implement this idea for the cyan contours in Fig. \ref{fi:DES_CS_seed1_200} and show that this brings the contours derived from a standard covariance estimate into consistency with those derived from the true covariance.

Downsides of this approach are a large increase of the uncertainties on cosmological parameters and the fact that one still needs at least as many realisations as data points in the data vector to even derive a precision matrix estimate. We now want to introduce an alternative method to estimate the precision matrix which is able to drastically decrease the offset of contours seen for the standard precision matrix estimator.

\section{Precision matrix expansion}
\label{sec:method}

Let us split the covariance matrix $\mathbf{C}$ into two contributions
\begin{equation}
\label{eq:split_of_covariance}
\mathbf{C} = \mathbf{A} + \mathbf{B}\ ,
\end{equation}
where for matrix $\mathbf{A}$ we have an accurate model (e.g. the shape-noise contributions to the covariance of cosmic shear correlation functions) and for $\mathbf{B}$ we have a model $\mathbf{B}_m$ which we know to be imperfect. We want to include this prior knowledge of the covariance matrix when estimating the precision matrix. Starting from 
\begin{equation}
\mathbf{C} = \mathbf{M} + (\mathbf{B}-\mathbf{B}_m)\ ,
\end{equation}
where $\mathbf{M} = \mathbf{A} + \mathbf{B}_m$ is our model for the complete covariance matrix, we rewrite
\begin{equation}
\label{eq:cov_in_terms_of_model}
\mathbf{C} = \left(\mathbb{1} + \mathbf{X}\right)\ \mathbf{M}\ ,
\end{equation}
where $\mathbb{1}$ is the identity matrix and we have defined
\begin{equation}
\mathbf{X} :=  (\mathbf{B}-\mathbf{B}_m)\ \mathbf{M}^{-1}\ .
\end{equation}
The precision matrix $\mathbf{\Psi} = \mathbf{C}^{-1}$ can then be expressed as the following power series in $\mathbf{X}$:
\begin{eqnarray}
\label{eq:power_series}
\mathbf{\Psi} &=& \mathbf{M}^{-1}\left(\sum_{k = 0}^\infty (-1)^k \mathbf{X}^k\right) \nonumber \\
&=& \mathbf{M}^{-1}\left(\mathbb{1} - \mathbf{X} + \mathbf{X}^2 + \mathcal{O}\left[\mathbf{X}^3\right] \right)\ .
\end{eqnarray}
We will call this series the \emph{precision matrix expansion} (PME). In appendix \ref{app:appendix_convergence} we show that it converges under a wide range of conditions. There we also demonstrate that the series yields at each order a symmetric approximation of $\mathbf{\Psi}$ and that at second order it is always positive definite (at each order if the series converges).

\subsection{Estimating the expansion of $\mathbf{\Psi}$}

Suppose we have an estimate $\smash{\mathbf{\hat B}}$ of the matrix $\smash{\mathbf{B}}$ from a number of N-body simulations. This especially assumes that all covariance contributions included in $\smash{\mathbf{A}}$ can be turned off in the simulations (i.e. for cosmic shear covariances $\smash{\mathbf{A}}$ could consist of shape-noise contributions which can be set to zero in simulations). We want to use $\smash{\mathbf{\hat B}}$ to construct unbiased estimators for the first order and second order term of the series in Eq. \ref{eq:power_series}.

Our assumptions state that $\smash{\mathbf{\hat B}}$ is drawn from a Wishart distribution with expectation value $\mathbf{B}$. In this case also $\smash{\mathbf{M}^{-1}\mathbf{\hat B}\mathbf{M}^{-1}}$ is Wishart distributed but with the expectation value $\smash{\mathbf{M}^{-1}\mathbf{B}\mathbf{M}^{-1}}$. Hence an unbiased estimator for the first order PME is given by
\begin{equation}
\label{eq:first_order_precision_estimator}
\mathbf{\hat \Psi}_{1\mathrm{st}} =\ \ \mathbf{M}^{-1}  - \mathbf{M}^{-1} \left(\mathbf{\hat B} - \mathbf{B}_m \right) \mathbf{M}^{-1}\ .
\end{equation}
Note that this does not involve the inversion of an estimated matrix. According to \citet{Taylor2013} the standard deviation of diagonal elements of an inverse-Wishart distributed matrix is proportional to $1/\sqrt{N_s - N_d - 4}$ while for Wishart distributed matrices it is only proportional to $1/\sqrt{N_s - 1}$. Hence, avoiding the occurence of an inverted matrix estimate greatly reduces the estimation noise.

The second order term involves squares of Wishart matrices. Using the results of \citet{Letac2004} on invariant moments of the Wishart distribution (cf. appendix \ref{app:LetacMassam2004}) it is still possible to construct an unbiased estimator for the second order PME as
\begin{eqnarray}
\label{eq:second_order_precision_estimator}
\mathbf{\hat \Psi}_{2\mathrm{nd}} &=&\ \ \mathbf{M}^{-1} + \mathbf{M}^{-1} \mathbf{B}_m \mathbf{M}^{-1}  \mathbf{B}_m \mathbf{M}^{-1} \nonumber \\
&& - \mathbf{M}^{-1} \left(\mathbf{\hat B} - \mathbf{B}_m \right) \mathbf{M}^{-1}  \nonumber \\
&& - \mathbf{M}^{-1} \mathbf{\hat B} \mathbf{M}^{-1} \mathbf{B}_m \mathbf{M}^{-1} \nonumber \\
&& - \mathbf{M}^{-1} \mathbf{B}_m \mathbf{M}^{-1} \mathbf{\hat B} \mathbf{M}^{-1} \nonumber \\
&& + \mathbf{M}^{-1} \frac{\nu^2 \mathbf{\hat B} \mathbf{M}^{-1} \mathbf{\hat B} - \nu \mathbf{\hat B}\ \mathrm{tr}\left( \mathbf{M}^{-1} \mathbf{\hat B}\right)} {\nu^2 + \nu - 2} \mathbf{M}^{-1}\ .
\end{eqnarray}
The estimator in Eq. \ref{eq:second_order_precision_estimator} is the key result of our paper. It has two advantages over the Anderson-Hartlap corrected standard estimator. First, it only requires matrix multiplications. As a consequence, it can even be used if $N_s \leq N_d$. Second, it only needs an estimate of $\mathbf{B}$ instead of the whole covariance $\mathbf{C}$, i.e. it allows to incorporate apriori knowledge on the covariance in the form of $\mathbf{M}$ (and $\mathbf{A}$).

In the next section we demonstrate that this significantly eases the requirement of $N_s - N_d \gg N_d - N_p$. Hence, in a likelihood analysis the noise in $\smash{\mathbf{\hat \Psi}_{2\mathrm{nd}}}$ becomes negligible for a much smaller number of N-body simulations than required by the standard precision matrix estimator. In appendix \ref{app:appendix_convergence} we also show that the bias in parameter constraints which arises from cutting the power series in Eq. \ref{eq:power_series} after a finite number of terms is negligible even for very strong deviations of our covariance model $\mathbf{M}$ from the N-body covariance $\mathbf{C}$.

\section{Examples: parameter errors for LSST weak lensing and DES weak lensing and multi-probe analyses}
\label{sec:examples}

\begin{figure}
  \includegraphics[width=8.5cm]{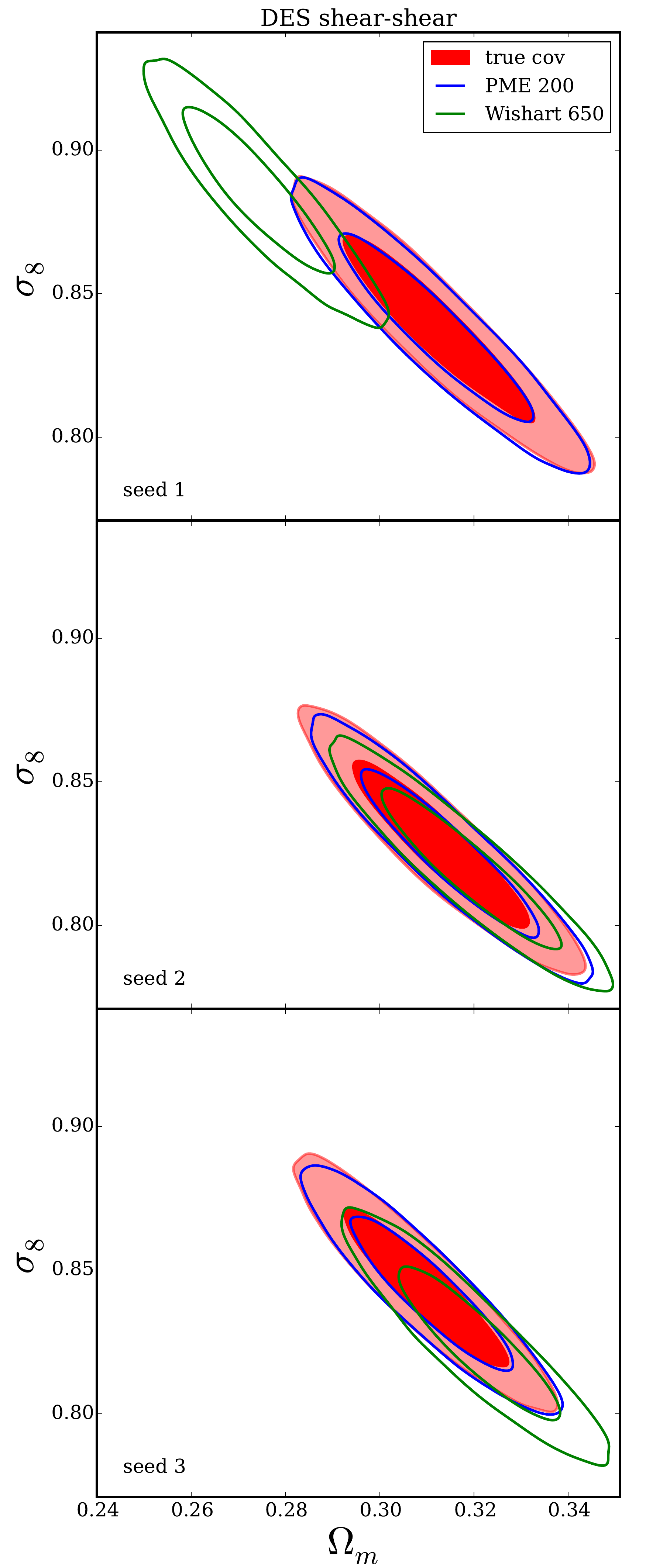}
   \caption{Contours in the $\Omega_m$-$\sigma_8$ plane obtained from realizations of our DES-like weak lensing data vector after marginalizing over all other parameters. For each random seed also new Wishart realisations $\mathbf{\hat B}$ and $\mathbf{\hat C}$ of the matrices $\mathbf{B}$ and $\mathbf{C}$ were drawn in order to simulate new realisations of the second order PME estimator and the standard precision matrix estimator. $N_s = 200$ simulations where assumed for the estimation of the PME while $N_s = N_d + 200 = 650$ simulations where assumed for the standard estimator.}
  \label{fi:DES_CS_200}
\end{figure}

\begin{figure}
  \includegraphics[width=8.5cm]{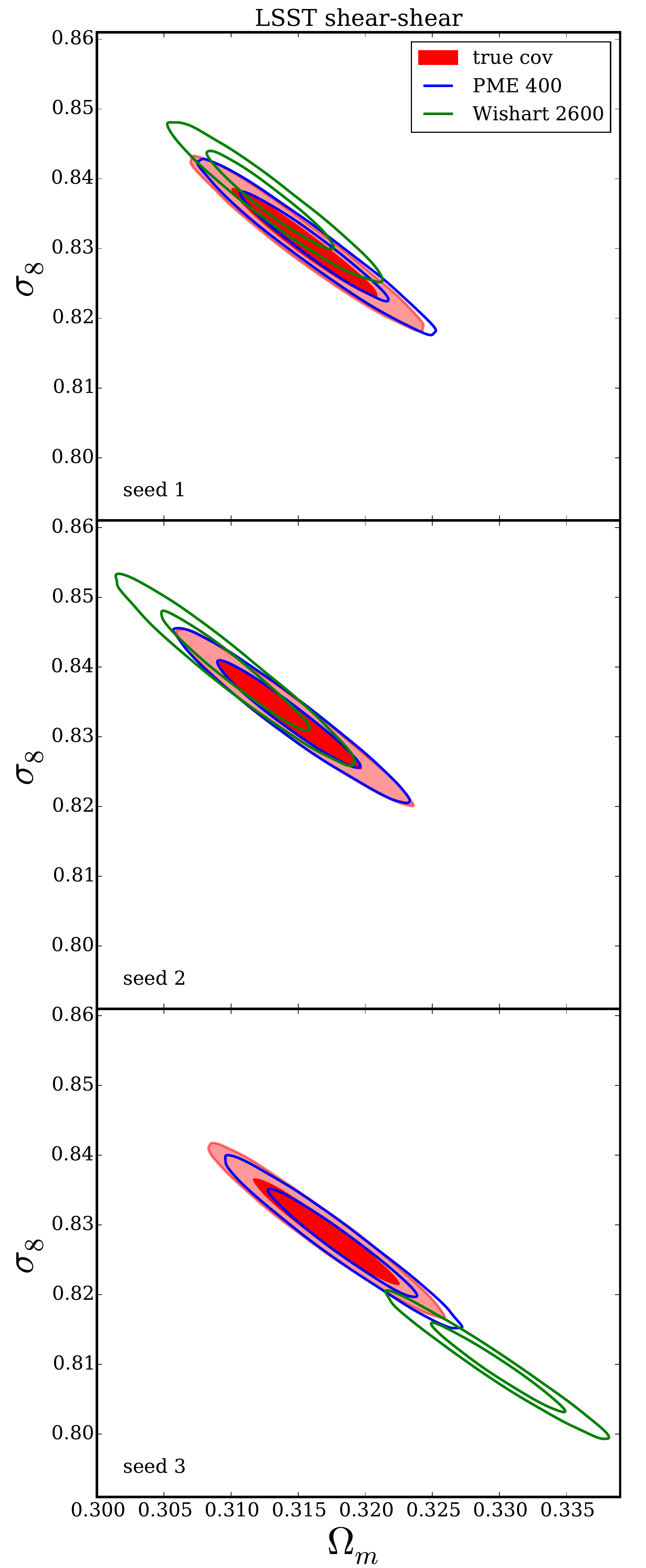}
   \caption{Same as Fig. \ref{fi:DES_CS_200} but for the LSST-like weak lensing data vector. $N_s = 400$ simulations where assumed for the estimation of the PME while $N_s = N_d + 400 = 2600$ simulations where assumed for the standard estimator.}
  \label{fi:LSST_400}
\end{figure}

We investigate the performance of our method in the context of ongoing and future surveys using DES and LSST as specific examples. These surveys differ in terms of survey area, galaxy number density, and redshift distribution and have different demands on the precision matrix. For DES we consider summary statistics in real space, i.e. auto- and cross-correlation functions of galaxy shear and position, for LSST we consider the corresponding Fourier quantities of a shear-shear only data vector. A summary of the scenarios considered is given in Table \ref{tab:scenarios} and a more detailed description of the considered data vectors is given in appendix \ref{app:data_vector}.

\begin{table}
\begin{center}
\begin{tabular}{ c || c | c | c | c  | c  } 
setup & survey &  lens bins & source bins & $N_{\mathrm{data}}$ & data \\ 
\hline
Ia & DES & 0  & 5 & 450 & real space \\
Ib & DES & 3  & 5 & 630 & real space \\
II  & LSST & 0 & 10 & 2200 & Fourier \\
\end{tabular}
\end{center}
\caption{Number of tomographic bins, total number of data points and type of data vector for the different setups used to test the performance of precision matrix expansion.}
 \label{tab:scenarios}
\end{table}

In order to test the performance of PME we set up mock experiments where we assume the true covariance matrix of each survey to be the analytic halo-model covariance described in \citet{Krause2016}. This model divides the covariance into three contributions: a noise-only part that consists of shape- and shot-noise contributions, $\mathbf{C}^{nn}$, a contribution from the cosmic variance of the signal, $\mathbf{C}^{ss, \mathrm{halo}}$, and a mixed term including noise and signal contributions, $\mathbf{C}^{sn}$. For shear-shear only covariances we set
\begin{equation}
\mathbf{A} = \mathbf{C}^{nn} + \mathbf{C}^{sn}
\end{equation}
and
\begin{equation}
\mathbf{B} = \mathbf{C}^{ss, \mathrm{halo}}\ .
\end{equation}
The shape-noise contributions to the covariance can be modelled reliably since the ellipticity dispersion can be measured from the data itself and since the mixed term $\mathbf{C}^{sn}$ involves only the modeling of two-point statistics of the shear field. The $\mathbf{B}$ term comprises the more complex 4-point statistics of the shear field, which can be estimated from simulations by turning off shape-noise. This is more complicated for galaxy clustering where shot-noise is included in the covariance matrix (cf. \ref{sec:DES_MP_noisy_PME}).

In order to simulate a situation where our covariance model $\mathbf{M} = \mathbf{A} + \mathbf{B}_m$ deviates from the true covariance we degrade it as
\begin{equation}
\label{eq:definition_of_pretend_model}
\mathbf{B}_m = \alpha\mathbf{C}^{ss, \mathrm{Gauss}} + \beta\left( \mathbf{C}^{ss, \mathrm{halo}} - \mathbf{C}^{ss, \mathrm{Gauss}}\right)
\end{equation}  
where $\smash{\mathbf{C}^{ss, \mathrm{Gauss}}}$ contains only the parts of the cosmic variance that are also present in a Gaussian covariance model. Hence, we allow the Gaussian and non-Gaussian cosmic variance parts to be over- or underestimated by a constant multiplicative factor. If not stated differently in this section we will use $\alpha = 1.0$ and $\beta = 0.5$. In appendix \ref{app:appendix_convergence} we explore a wider range of rescalings and also consider more complex deformations of our fiducial covariance to show that the PME remains robust under more complicated deviations of $\mathbf{M}$ from the true covariance matrix. All simulated likelihood analyses in this paper are computed using the \verb|CosmoLike| cosmology package \citep{ Krause2016, Eifler2014}.

\subsection{Performance for DES weak lensing data vector}

We now carry out mock likelihood analyses for DES and LSST weak lensing data vectors, varying the parameters $\Omega_m$, $\sigma_8$, $w_0$ and $w_a$. Our fiducial values for these parameters are
\begin{eqnarray}
&(\Omega_m, \sigma_8, w_0, w_a) =&\nonumber \\ &(0.3156,\ 0.831,\ -1,\ 0)\ .&
\end{eqnarray}
We start by drawing random Gaussian realisations of our fiducial data vectors according to a covariance given by the halo model. For each realisation we also draw new Wishart realisations $\mathbf{\hat B}$ and $\mathbf{\hat C}$ of cosmic variance and total covariance to compute the PME estimate $\smash{\mathbf{\hat \Psi}_{\mathrm{2nd}}}$ and the standard estimator $\smash{\mathbf{\hat \Psi}}$. In practice, this is done by drawing additional realisations $\smash{\boldsymbol{\hat\xi}_i}$\ , $i = 1\ \dots\ N_s$, of our fiducial data vector from a multivariate Gaussian distribution whose covariance is $\mathbf{B}$ respectively $\mathbf{C}$. These realisations represent measurements from N-body simulations and inserting them into Eq. \ref{eq:cov_estimator} generates the desired Wishart realisations $\mathbf{\hat B}$ and $\mathbf{\hat C}$ of the two matrices.

Using \verb|CosmoLike| we then run likelihood chains to infer a posterior distribution for our parameters using $\smash{\mathbf{\hat \Psi}_{\mathrm{2nd}}}$, $\smash{\mathbf{\hat \Psi}}$ and the true precision matrix $\smash{\mathbf{C}^{-1}}$. When computing the likelihood from $\smash{\mathbf{C}^{-1}}$ and $\smash{\mathbf{\hat \Psi}_{\mathrm{2nd}}}$ we simply use standard ansatz given in Eq. \ref{eq:standard_likelihood}. When deriving contours from the Wishart realisation $\smash{\mathbf{\hat C}}$ we furthermore compute the parameter likelihood as
\begin{equation}
p(\boldsymbol{\pi}|\boldsymbol{\hat \xi}) \sim \left[1 + \frac{\left(\boldsymbol{\hat \xi} - \boldsymbol{\xi}[\boldsymbol{\pi}] \right)^T \mathbf{\hat C}^{-1} \left(\boldsymbol{\hat \xi} - \boldsymbol{\xi}[\boldsymbol{\pi}] \right)}{N_s -1}\right]^{-N_s / 2}
\end{equation}
which SH16a have shown to be a more accurate than using the Kaufman-Hartlap correction and the standard Gaussian likelihood. We however found only small differences to using the standard likelihood ansatz, which is due to the fact that in all cases considered in this paper $N_s - N_d \gg N_p$.

In Fig. \ref{fi:DES_CS_200} and \ref{fi:LSST_400} we show see the resulting 1$\sigma$ and 2$\sigma$ contours in the $\Omega_m$-$\sigma_8$ plane (after marginalizing over the other parameters) for 3 different random draws of data vector and Wishart matrices. For each realisation of the DES data vector we assumed that $N_s = 200$ simulations are available to estimate the PME and $N_s = N_d + 200 = 650$ simulations for the standard estimator. For each realisation of the LSST data vector we assumed $N_s = 400$ simulations for the PME and $N_s = N_d + 400 = 2600$ simulations for the standard estimator.

Even though in each case we assumed many more simulations for the standard estimator than for the PME, the PME is significantly better in reconstructing the contours from the true precision matrix. In particular we find that deviations from the true contours are much smaller than the corresponding 1$\sigma$ and 2$\sigma$ uncertainties of the parameters.

Next we generalize the findings in Figs. \ref{fi:DES_CS_200} and \ref{fi:LSST_400}. We generate $1000$ Wishart realisations of the matrices $\smash{\mathbf{\hat C}}$ and $\smash{\mathbf{\hat B}}$ for different assumptions on the number of available N-body simulations $N_s$. For each of the $1000$ sets of matrices we also generate $10$ realizations $\smash{\boldsymbol{\hat \xi}}$ of our fiducial data vector (i.e. overall $10\,000$ different realizations $\smash{\boldsymbol{\hat \xi}}$). Hence for each type of precision matrix estimate we perform overall $10\,000$ likelihood analyses. In each analysis we determine the best fit parameters $\boldsymbol{\hat \pi}_{\mathrm{ML}}$ and check whether our fiducial cosmology is outside the $68.3\%$ confidence contour around these parameters. In order to make this computationally feasible, we are now linearly approximating the calculations of \verb|CosmoLike| around our fiducial cosmology $\boldsymbol{\pi}_0$, i.e. we use
\begin{equation}
\boldsymbol{\xi}_{\mathrm{simple}}[\boldsymbol{\pi}] = \boldsymbol{\xi}_{\mathrm{exact}}[\boldsymbol{\pi}_0] + \sum_{i=1}^{N_p} (\pi_i - \pi_{0,i}) \frac{\partial \boldsymbol{\xi}_{\mathrm{exact}}}{\partial \pi_i }[\boldsymbol{\pi}_0]\ .
\end{equation}
This allows us to analytically determine the maximum likelihood parameters and the $68.3\%$ confidence contours in each likelihood analysis. It is also the situation where a Fisher-matrix formalism and hence the derivations of DS13 hold exactly.

\begin{figure*}
\includegraphics[width=17.0cm]{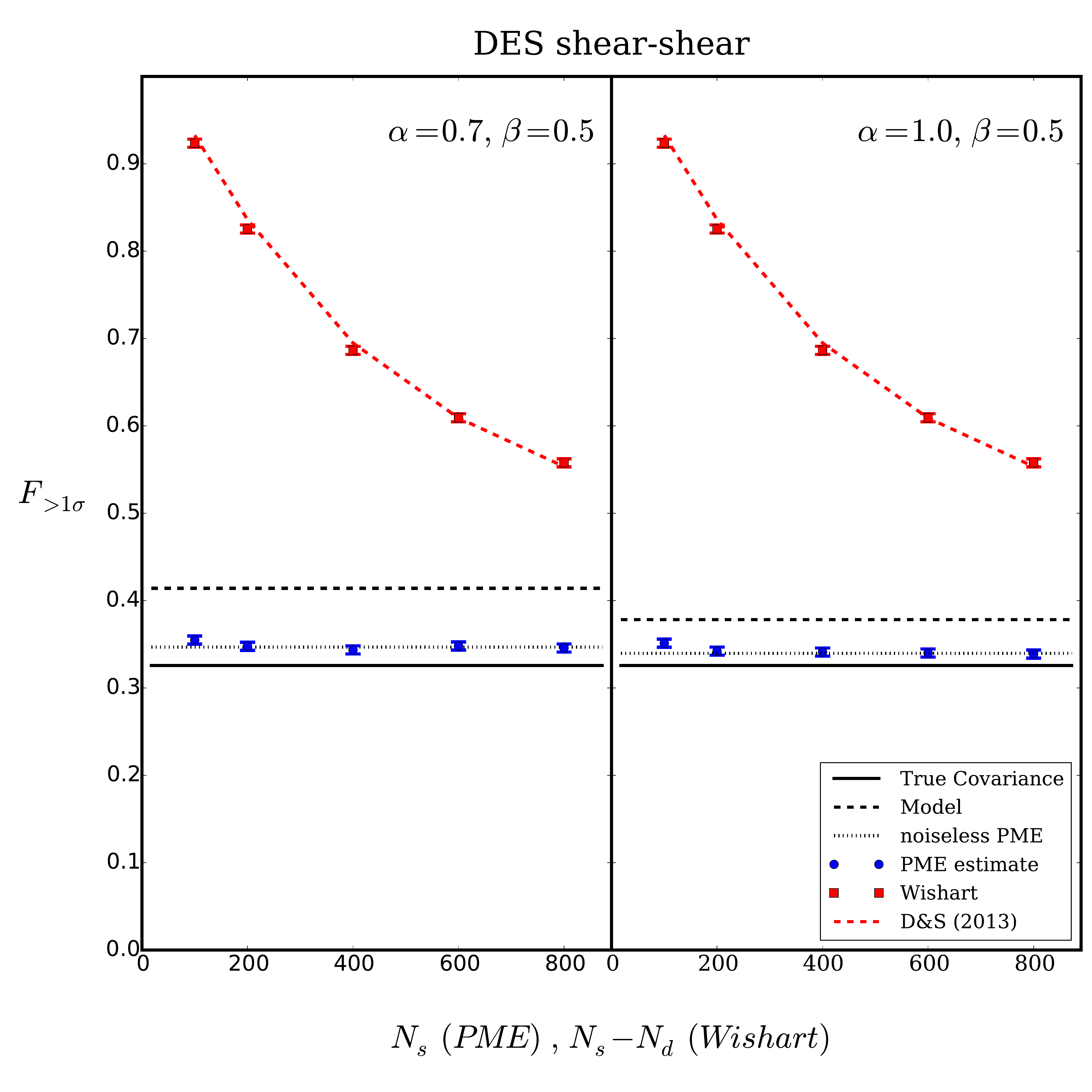}
   \caption{The figure compares $F_{>1\sigma}$, the number of times that our fiducial cosmology was considered outside the $68.3\%$ confidence contour in our simulated likelihood analyses when using different precision matrix estimates for computing the posterior parameter likelihood. In order to carry a sufficient number of mock analyses, we simplified our modeling of the data vector by linearly approximating the full computation around our fiducial cosmology. For the DES-like weak lensing data vector we varied the four parameters $(\Omega_m, \sigma_8, w_0, w_a)$.}
  \label{fi:F1sigma_DES_shear-shear}
\end{figure*}

We define $\smash{F_{>1\sigma}}$ as the fraction of times that our fiducial cosmology is outside of the $68.3\%$ confidence contour around the best fit parameters and we use it as a metric for comparing the different precision matrix estimators. In Fig. \ref{fi:F1sigma_DES_shear-shear} we show this fraction for all different types of precision matrices introduced before. The solid, dashed, and dotted lines show the fractions achieved when using the noise-less matrices $\smash{\mathbf{C}^{-1}}$, $\smash{\mathbf{M}^{-1}}$ and $\smash{\mathbf{\Psi}_{\mathrm{2nd}}}$. Especially, the noise-less matrix $\smash{\mathbf{\Psi}_{\mathrm{2nd}}}$ would be the PME-estimator in the limit of infinitely many simulations and $\smash{\mathbf{C}^{-1}}$ would be the standard estimator in the same limit. The red and blue dots show the fraction achieved when using the noisy precision matrix estimates $\smash{\mathbf{\hat \Psi}}$ and $\smash{\mathbf{\hat \Psi}_{\mathrm{2nd}}}$.

As expected, $\smash{F_{>1\sigma}}$ is very close to $32\%$ when using the true covariance $\mathbf{C}$ in the likelihood analyses. For the deformed halo model covariance $\mathbf{M}$ we assumed the two cases $\alpha = 0.7$, $\beta = 0.5$ (left panel) and $\alpha = 1.0$ and $\beta = 0.5$ (right panel). For $\alpha = 0.7$ and $\beta = 0.5$ our fiducial cosmology is regarded as outside the $68.3\%$ contour in more than $40\%$ of the cases. For both choices of $\mathbf{M}$ the noise-free PME significantly corrects that fraction towards the optimal value of $\sim32\%$. Especially promising is that the PME estimate performs very similar to the noise-free PME. If 200 simulations are available to estimate the PME, it essentially converges to its best possible performance. And even if only $100$ simulations are available to estimate the PME, its value of $\smash{F_{>1\sigma}}$ comes closer to $32\%$ than when using $\mathbf{M}$ to derive the contours.

When inferring the likelihood from the standard precision matrix estimator $\smash{F_{>1\sigma}}$ is greater than $50\%$ even if we allow $N_s = N_d + 800$ simulations for the covariance estimation, which corresponds to $1250$ simulations. This is due to the additional variance of $\boldsymbol{\hat \pi}_{\mathrm{ML}}$ caused by the noise of the precision matrix (cf. Eq. \ref{eq:best_fit}). Using the results of DS13 we can derive predictions for this effect (cf. appendix \ref{app:Dodelson2013}). As can be seen from the red dashed lines in Fig. \ref{fi:F1sigma_DES_shear-shear} these predictions agree well with what we find in our simulated likelihood analyses. Extrapolating the results of DS13 to higher values of $N_s$ we can also estimate, how many simulations would be required for the standard precision matrix estimator in order to achieve the same value of $\smash{F_{>1\sigma}}$ as the second order PME. For the left panel of Fig. \ref{fi:F1sigma_DES_shear-shear} we find that it would take $\sim 8000$ simulations for the standard estimator to get as close to $\smash{F_{>1\sigma} = 32\%}$ as the PME with only $200$ simulations. This statement however depends on the model covariance $\mathbf{M}$ since it determines how well the PME has converged after its second order.

An $\mathbf{M}$-independent way of comparing standard estimator and PME estimator is to see how many simulations it takes each to have $F_{>1\sigma}$ within $1\%$ of their best possible performance. It would take the standard estimator $\sim 24\,000$ simulations to be within $1\%$ of $F_{>1\sigma} = 32\%$. The PME estimator is well within $1\%$ of its best possible performance for only 200 simulations.

Note that with the results of DS13 one can in principle correct a likelihood analysis for the additional variance caused by the standard precision matrix estimator. This would result in a decreased constraining power of the analysis and it would hence be the main benefit of the PME to prevent this loss.

\subsubsection{Larger covariance matrices: LSST weak lensing data vector}
\label{sec:LSST_noisy_PME}

\begin{figure}
  \includegraphics[width=8.5cm]{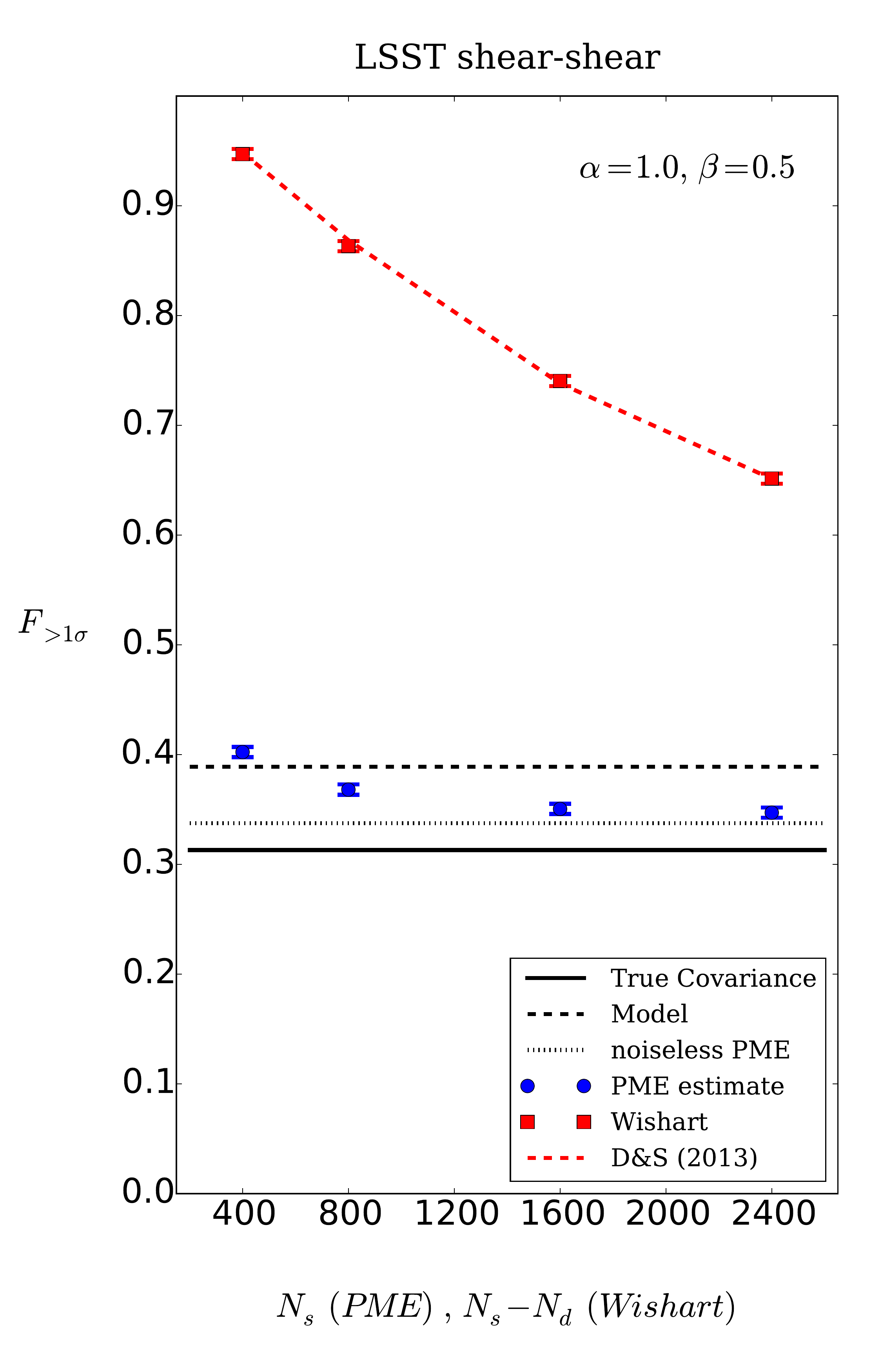}
   \caption{Same as Fig. \ref{fi:F1sigma_DES_shear-shear} but for the LSST-like weak lensing data vector.}
  \label{fi:F1sigma_LSST}
\end{figure}

We repeat the above analysis for the LSST-like weak lensing data vector. Fig. \ref{fi:F1sigma_LSST} shows the fractions $F_{>1\sigma}$ obtained from PME and standard precision matrix estimator. The PME estimator now requires $\sim 2400$ simulations to be less than $1\%$ away from its best possible performance. As before, this statement does not include any additional biases between PME and true precision matrix that might arise from the biased model matrix $\mathbf{M}$ used to carry out the matrix expansion. The standard precision matrix estimator would need $N_s > 115\,000$ simulations to be less than $1\%$ away from its best possible performance.

\begin{figure}
  \includegraphics[width=8.5cm]{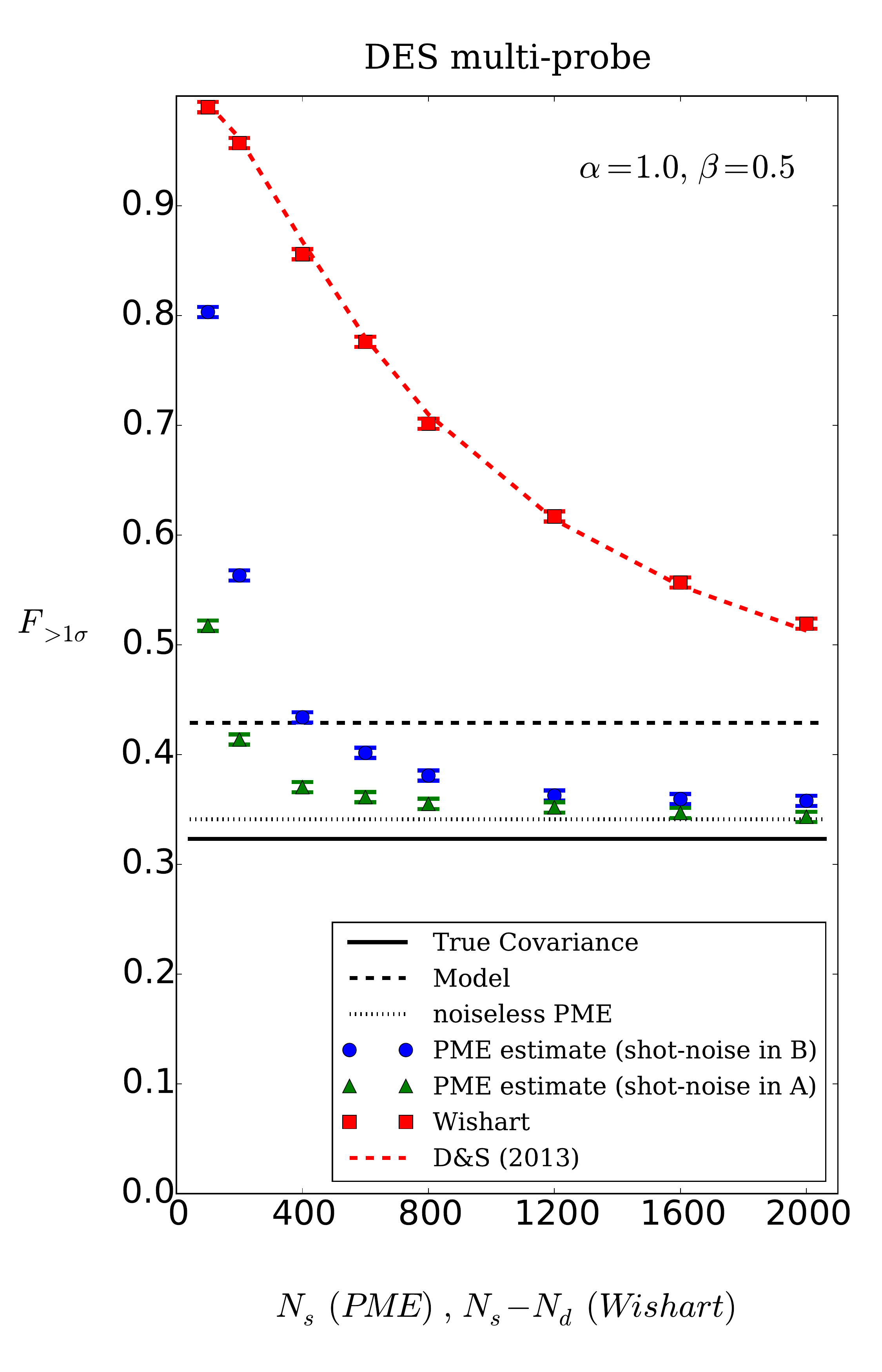}
   \caption{Same as Fig. \ref{fi:F1sigma_DES_shear-shear} but for the DES multi-probe data vector. For this case the 7 parameters $(\Omega_m, \sigma_8, w_0, w_a, b_1, b_2, b_3)$ were varied in each likelihood analysis. Green points assume that cosmic variance can be estimated from simulations without shot-noise. This would significantly improve the performance of PME for low numbers of available simulations.}
  \label{fi:F1sigma_DES_multi-probe}
\end{figure}

\subsubsection{Defining $\mathbf{A}$ and $\mathbf{B}$ for multi-probe covariances}
\label{sec:DES_MP_noisy_PME}

We now repeat the analysis of Fig. \ref{fi:F1sigma_DES_shear-shear} for a DES-like multi-probe data vector. This vector includes contributions from galaxy clustering and galaxy-galaxy lensing, which introduces shot-noise terms to the covariance. These shot-noise contributions are in principle well understood theoretically and include, similar to the cosmic shear case, at most two-point statistics of the cosmic density field. Hence, one could absorb them into the matrix $\mathbf{A}$ (cf. Eq. \ref{eq:split_of_covariance}) and use N-body simulations only for the remaining part of the covariance - i.e. to define $\mathbf{B}$ as only the cosmic variance. This is however difficult since most N-body simulations provide only simulated galaxy catalogues that are affected by shot noise themselves, which makes it impossible to independently estimate the cosmic variance. If however all shot-noise contributions are included in $\mathbf{B}$ when defining and estimating the PME, then the estimator $\smash{\mathbf{\hat \Psi}_{2\mathrm{nd}}}$ will have a higher variance in many of its elements. Hence, the additional scatter of best fit parameters due to a noisy precision matrix might not be negligible anymore.

In Fig. \ref{fi:F1sigma_DES_multi-probe} we compare the fractions $F_{>1\sigma}$ obtained from different estimates of the precision matrix in our simulated likelihood analyses - this time for the DES multi-probe data vector. In each likelihood analysis we now vary $7$ parameters, since for each lens bin we include a galaxy bias parameter in our model. The fiducial bias values are
\begin{equation}
(b_1,\ b_2,\ b_3) = (1.35,\ 1.50,\ 1.65)
\end{equation}
in order of increasing redshift. Fig. \ref{fi:F1sigma_DES_multi-probe} shows the results obtained for each of the mentioned options of defining $\mathbf{B}$. It is clear that the noisy PME approaches its best possible performance already for a smaller number of simulations if the cosmic variance can be estimated directly. In practice this would however require density maps in thin redshift slices for each simulation in order to measure the correlation functions of the projected density fields without shot-noise.

\begin{figure}
  \includegraphics[width=8.5cm]{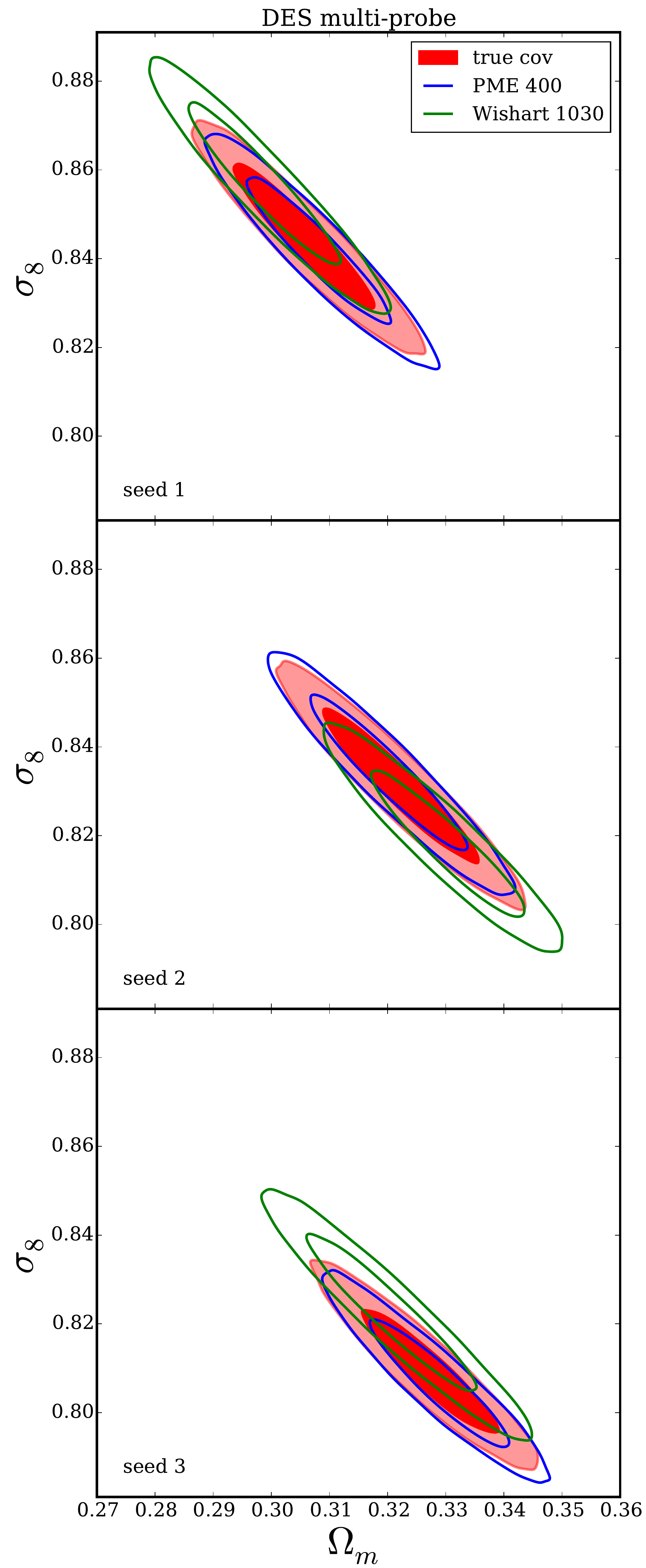}
   \caption{Same as Fig. \ref{fi:DES_CS_200} but for the multi-probe data vector. $N_s = 400$ simulations where assumed for the estimation of the PME while $N_s = N_d + 400 = 1030$ simulations where assumed for the standard estimator. Even with fewer simulations the PME is much better in reconstructing the contours from the true precision matrix. However, below $N_s = 1600$ a significant offset of the contours persists.}
  \label{fi:DES_MP_400}
\end{figure}

Assuming one can directly measure the cosmic variance from simulations we again want to asses how many simulations are required for the standard precision matrix estimator and the PME estimator to be within $1\%$ of their best possible performance. Extrapolating the results of DS13 we find that it would take the standard estimator $\sim 44\,000$ simulations to be within within $1\%$ of $F_{>1\sigma} = 32\%$. The PME estimator is within $1\%$ of its best possible performance for 1600 simulations. For $N_s = 2000$ the performance of the PME becomes almost solely restricted by the deviation between $\mathbf{M}$ and $\mathbf{C}$ in our mock experiment. However, below $N_s = 1600$ there seems to be significant additional scatter of the best fitting parameters due to the noise of the PME estimate. We demonstrate this in Fig. \ref{fi:DES_MP_400} for $N_s = 400$. Regardless of how $\mathbf{B}$ is defined, we can nevertheless conclude that also for multi-probe covariances the PME poses a vast improvement over the standard precision matrix estimator.

\section{Conclusions}
\label{sec:conclusions}

It was the starting point of our analysis to find a method for using a priori knowlegde about the covariance matrix when estimating the precision matrix from simulations. This requires finding an equivalent of the Kaufman-Hartlap correction when only parts of the covariance are estimated. Using the results of \citet{Letac2004} we partly solved this task by calculating an expansion of the precision matrix and showing how the leading terms of this expansion can be estimated from simulations. Our method enables the use of preexisting knowledge on the covariance structure to improve the convergence of the PME and to reduce the noise in its estimation. It also has the advantage that the relative uncertainties of the elements of the PME estimate scale with the number of available simulations $N_s$ as $\sim 1/\sqrt{N_s - 1}$, which is typically much smaller than the uncertainties of the standard precision matrix estimator. The latter also depend on the number of data points $N_d$ and scale as $\sim 1/\sqrt{N_s - N_d - 4}$.

We demonstrated that the PME converges even for drastic deviations between the model covariance and the N-body covariance and we also showed that it provides a much less noisy estimate of the parameter likelihood compared to estimating the precision matrix in the standard way. For a DES weak lensing data vector $N_s \gtrsim 8000$ simulations would be required for the standard estimator to reconstruct the likelihood similarly well as the PME with only $N_s = 200$ - even if the model covariance heavily underestimates Gaussian and non-Gaussian covariance parts. If we assume more realistic deviations between model and N-body covariance, up to $24\,000$ simulations would be needed for the standard estimator to reconstruct the 1$\sigma$ quantile of the parameter distribution at the same precision as the PME with only 200 simulations. For an LSST-like weak lensing data vector with $N_d = 2200$ we found that up to $115\,000$ simulations would be required for the standard estimator to reconstruct the 1$\sigma$ quantile as well as the PME with only 2400 simulations. It should however be stressed that these statements depend on the quality of the model covariance $\mathbf{M}$ that was used to compute the PME.

Additional complications arise when galaxy clustering correlation functions are included in the data vector. A performance similar to the weak lensing case can still be achieved if one manages to estimate the cosmic variance of the correlation functions directly, i.e. without shot-noise. For this case, we find that a DES-like multi-probe data vector requires up to $44\,000$ simulations for the standard precision matrix estimator to reconstruct the 1$\sigma$ quantile of the parameter distribution as well as the PME with 1600 simulations.

One aspect that should be addressed in future work, is to find a priori criteria for the convergence of the PME. In appendix \ref{app:appendix_convergence} we demonstrate that it converges for very strong deformations of the halo-model covariance, but one can not be certain whether and how fast it will converge for all possible data vectors and covariance models. As we show in appendix \ref{app:appendix_convergence}, situations where the PME does not converge can at least be identified a posteriori by a comparison of the first order and second order expansion. A strong oscillation of likelihood contours derived from the first and second order PME indicate a significant deviation of model and N-body covariance. This way, the PME provides a clear criterion for testing covariance models with simulations -- even when the number of available simulations is small.

\newpage

\section*{Acknowledgments}

This work was supported by SFB-Transregio 33 `The Dark Universe'  by  the  Deutsche  Forschungsgemeinschaft  (DFG). We also acknowledge the support by the DFG Cluster of Excellence "Origin and Structure of the Universe". The simulations have been carried out on the computing facilities of the Computational Center for Particle and Astrophysics (C2PAP). This research is partially supported by NASA ROSES ATP 16-ATP16-0084 grant. Part of the research was carried out at the Jet Propulsion Laboratory, California Institute of Technology, under a contract with the National Aeronautics and Space Administration. 

This paper has gone through internal review by the DES collaboration. We want to thank Eric Baxter, Gary Bernstein, Scott Dodelson, Franz Elsner, Juan Garcia-Bellido, John Peacock, Stella Seitz and the DES multi-probe pipelines group for very helpful comments and discussions on this project.

Funding for the DES Projects has been provided by the U.S. Department of Energy, the U.S. National Science Foundation, the Ministry of Science and Education of Spain, 
the Science and Technology Facilities Council of the United Kingdom, the Higher Education Funding Council for England, the National Center for Supercomputing 
Applications at the University of Illinois at Urbana-Champaign, the Kavli Institute of Cosmological Physics at the University of Chicago, 
the Center for Cosmology and Astro-Particle Physics at the Ohio State University,
the Mitchell Institute for Fundamental Physics and Astronomy at Texas A\&M University, Financiadora de Estudos e Projetos, 
Funda{\c c}{\~a}o Carlos Chagas Filho de Amparo {\`a} Pesquisa do Estado do Rio de Janeiro, Conselho Nacional de Desenvolvimento Cient{\'i}fico e Tecnol{\'o}gico and 
the Minist{\'e}rio da Ci{\^e}ncia, Tecnologia e Inova{\c c}{\~a}o, the Deutsche Forschungsgemeinschaft and the Collaborating Institutions in the Dark Energy Survey. 
The DES data management system is supported by the National Science Foundation under Grant Number AST-1138766.

The Collaborating Institutions are Argonne National Laboratory, the University of California at Santa Cruz, the University of Cambridge, Centro de Investigaciones En{\'e}rgeticas, 
Medioambientales y Tecnol{\'o}gicas-Madrid, the University of Chicago, University College London, the DES-Brazil Consortium, the University of Edinburgh, 
the Eidgen{\"o}ssische Technische Hochschule (ETH) Z{\"u}rich, 
Fermi National Accelerator Laboratory, the University of Illinois at Urbana-Champaign, the Institut de Ci{\`e}ncies de l'Espai (IEEC/CSIC), 
the Institut de F{\'i}sica d'Altes Energies, Lawrence Berkeley National Laboratory, the Ludwig-Maximilians Universit{\"a}t M{\"u}nchen and the associated Excellence Cluster Universe, 
the University of Michigan, the National Optical Astronomy Observatory, the University of Nottingham, The Ohio State University, the University of Pennsylvania, the University of Portsmouth, 
SLAC National Accelerator Laboratory, Stanford University, the University of Sussex, and Texas A\&M University.

The DES participants from Spanish institutions are partially supported by MINECO under grants AYA2012-39559, ESP2013-48274, FPA2013-47986, and Centro de Excelencia Severo Ochoa SEV-2012-0234.
Research leading to these results has received funding from the European Research Council under the European Union’s Seventh Framework Programme (FP7/2007-2013) including ERC grant agreements 
 240672, 291329, and 306478.


\bibliographystyle{mnras}
\bibliography{precision_matrix_expansion}

\appendix

\section{Influence of noisy covariance estimates on the scatter of best fitting cosmological parameters}
\label{app:Dodelson2013}

Using a noisy precision matrix estimate to determine the best fitting cosmological parameters
\begin{equation}
\boldsymbol{\hat \pi}_{\mathrm{ML}} = \underset{\boldsymbol{\pi}}{\min} \left\lbrace\left(\boldsymbol{\hat \xi} - \boldsymbol{\xi}[\boldsymbol{\pi}] \right)^T \mathbf{\Psi} \left(\boldsymbol{\hat \xi} - \boldsymbol{\xi}[\boldsymbol{\pi}] \right)\right\rbrace\ .
\end{equation}
leads to an additional scatter in these parameters. Especially, this additional noise is not accounted for by the width of contours generated from the precision matrix estimate. This effect has e.g. been described by \citet{Dodelson2013} who also derived a prediction for the additional noise assuming a Gaussian parameter likelihood. They find that the actual parameter covariance when using an inverse-Wishart realisation of the precision matrix is given by
\begin{equation}
\label{eq:DS2013}
\mathbf{C}_{\boldsymbol{\hat \pi}_{\mathrm{ML}}} = \mathbf{F}^{-1} \left(1 + \frac{(N_d - N_p)(N_s - N_d - 2)}{(N_s - N_d - 1)(N_s - N_d - 4)} \right)\ ,
\end{equation}
where $N_p$ is the number of considered parameters and $\mathbf{F}$ is the Fisher matrix computed from the true precision matrix. Hence, in the case of a Gaussian parameter likelihood, best fit parameters $\boldsymbol{\hat \pi}_{\mathrm{ML}}$ that are computed from a Wishart realisation of the covariance have also a Gaussian distribution but with a rescaled parameter covariance.

\section{Unbiased estimator of the square of a Wishart matrix}
\label{app:LetacMassam2004}

Let $\smash{\mathbf{\hat C}}$ be distributed according to a Wishart distribution with $\nu$ degrees of freedom and expectation value $\mathbf{C}$. Then
\begin{equation}
\left\langle \mathbf{\hat C}^2\right\rangle \neq \mathbf{C}^2\ .
\end{equation}
However, using the results of \citet{Letac2004} it is possible to devise an unbiased estimator of $\mathbf{C}^2$. It is given by
\begin{equation}
\widehat{\left( \mathbf{C}^2 \right)} = \frac{\nu^2 \mathbf{\hat C}^2 - \nu \mathbf{\hat C} \mathrm{tr} \mathbf{\hat C}}{\nu^2 + \nu - 2}\ ,
\end{equation}
where $\smash{\mathrm{tr} \mathbf{\hat C}}$ denotes the trace of $\smash{\mathbf{\hat C}}$. Using this formula, it is straight foreward to derive the estimator of the second order PME given in Eq. \ref{eq:second_order_precision_estimator}.

\section{General properties and convergence of the power series}
\label{app:appendix_convergence}

\subsection{General properties}

In order to derive some general properties of the PME series, let us slightly change the notation of Sec. \ref{sec:method}. First, let $\mathbf{M}^{1/2}$ be the unique symmetric and positive definite matrix such that
\begin{equation}
\mathbf{M}^{1/2} \mathbf{M}^{1/2} = \mathbf{M}\ .
\end{equation}
This matrix exists as long as our covariance model $\mathbf{M}$ is positive definite. Let us then re-define
\begin{equation}
\mathbf{X}  = \mathbf{M}^{-1/2}\left(\mathbf{B} - \mathbf{B}_m \right)\mathbf{M}^{-1/2}
\end{equation}
where $\mathbf{M}^{-1/2}$ is the inverse of $\mathbf{M}^{1/2}$, and $\mathbf{B}$ and $\mathbf{B}_m$ are the same as in Sec. \ref{sec:method}. The complete covariance can then be written as
\begin{equation}
\mathbf{C}  = \mathbf{M}^{1/2}\left(\mathbb{1} + \mathbf{X} \right)\mathbf{M}^{1/2}
\end{equation}
and the precision matrix expansion now reads
\begin{eqnarray}
\mathbf{\Psi} &=& \mathbf{M}^{-1/2}\left(\sum_{k = 0}^\infty (-1)^k \mathbf{X}^k\right)\mathbf{M}^{-1/2} \nonumber \\
&=& \mathbf{M}^{-1/2}\left(\mathbb{1} - \mathbf{X} + \mathbf{X}^2 + \mathcal{O}\left[\mathbf{X}^3\right] \right)\mathbf{M}^{-1/2}\ .
\end{eqnarray}
Since both $\mathbf{M}^{-1/2}$ and $\mathbf{X}$ are symmetric matrices, it is immediately clear that this gives a symmetric approximation of $\mathbf{\Psi}$ at each order of the power series. The series converges if and only if all eigenvalues of $\mathbf{X}$ fulfill
\begin{equation}
|\lambda_i| < 1\ ,\ i = 1\ ,\ ...\ ,\ N_d\ .
\end{equation}
In each eigendimension of $\mathbf{X}$ the series $\smash{\left(\mathbb{1} - \mathbf{X} + \mathbf{X}^2 + \mathcal{O}\left[\mathbf{X}^3\right] \right)}$ is simply the geometric series. For $|\lambda_i| < 1$ the value of this series is $ > 0$ at each finite order. At second order, the value of this series is $ > 0$ regardless of the values of $\lambda_i$. Hence, the second order PME is always positive definite.

\subsection{Special cases}

\subsubsection{Rescaling of the covariance}
\label{sec:rescaling_of_cov}
Let us investigate the convergence properties of the power series in Eq. \ref{eq:power_series} in a couple of special cases. We start by assuming that our model for the covariance matrix, $\mathbf{M}$, under- or overestimates the true covariance matrix by a constant factor $\alpha$, i.e.
\begin{equation}
\mathbf{M} = \alpha \mathbf{C}\ .
\end{equation}
In this case we have
\begin{eqnarray}
\mathbf{X} &=& \mathbf{M}^{-1/2} \left( \mathbf{C} - \alpha \mathbf{C}\right)\mathbf{M}^{-1/2}\nonumber \\
 &=& \frac{1-\alpha}{\alpha}\mathbf{C}^{-1/2}\mathbf{C}\mathbf{C}^{-1/2}\nonumber \\
  &=& \frac{1-\alpha}{\alpha}\mathbb{1}\ .
\end{eqnarray}
Hence, all eigenvalues of $\mathbf{X}$ are given by $\lambda = \frac{1-\alpha}{\alpha}$. This has absolute value smaller than $1$ for all $\alpha > 0.5$. This especially means that the series used to define the PME converges even if the model covariance overestimates the true covariance by an arbitrarily high overall factor. Since we cut Eq. \ref{eq:power_series} after the second order we must however look at how well the series is converged after that order. The relative error on each element of the precision matrix is given by
\begin{equation}
\label{eq:appendix_convergence}
\frac{\Psi_{ij} - \Psi_{2\mathrm{nd},ij}}{\Psi_{ij}} = \lambda^3 = \frac{(1-\alpha)^3}{\alpha^3}\ .
\end{equation}
This is $<10\%$ for $\alpha \in [0.69 , 1.86]$ and $<1\%$ for $\alpha \in [0.83, 1.27]$.

\subsubsection{Partial rescaling of the covariance}

\begin{figure*}
  \includegraphics[width=17.0cm]{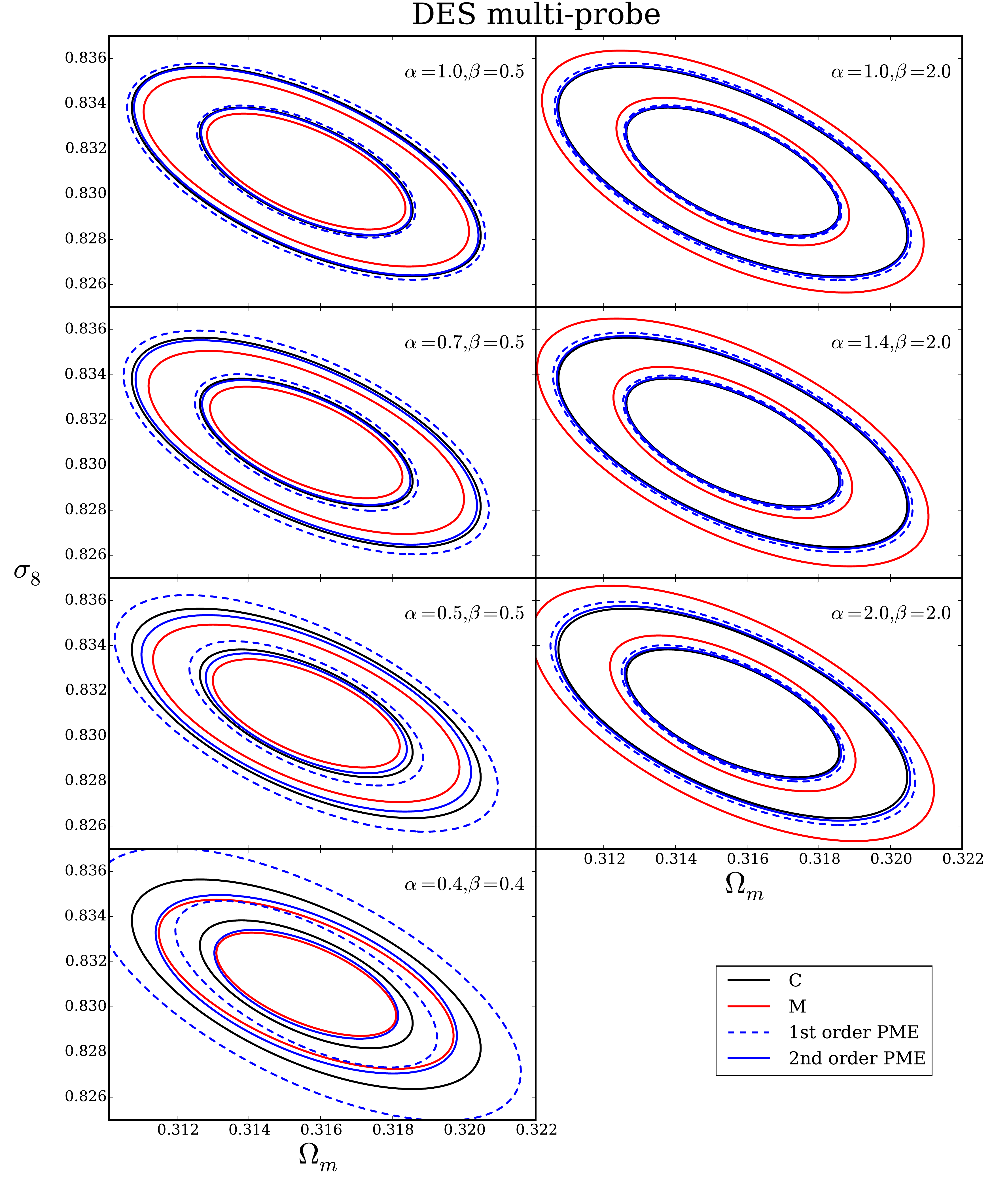}
   \caption{We show the $1\sigma $ and $2\sigma $ Fisher contours in the $\Omega_m$-$\sigma_8$ plane around our fiducial cosmology using the DES multi-probe data vector and keeping all other cosmological parameters fixed. For the black contours the Fisher matrix was derived from the fiducial covariance matrix $\mathbf{C}$ of our experiment -- the halo-model covariance. For the red contours we rescaled the Gaussian and non-Gaussian parts of the cosmic variance in $\mathbf{C}$ by constant factors $\alpha$ and $\beta$ to create our model covariance matrix $\mathbf{M}$ (cf. Eq. \ref{eq:definition_of_pretend_model}). The blue contours show the constraints derived from the 1st order PME (dashed lines) and 2nd order PME (solid lines) of $\mathbf{C}$ around $\mathbf{M}$. The PME manages to significantly correct the miss estimation of the Fisher matrix by the model precision matrix for most values of the rescaling factors. Only for $\alpha,\ \beta < 0.5$ the convergence of the PME seems to break down and a strong oscillation between 1st order and 2nd order correction occurs. We discuss this behavior in detail in appendix \ref{app:appendix_convergence} where we also study examples of more complicated deviations between $\mathbf{M}$ and $\mathbf{C}$.}
  \label{fi:cons_multi}
\end{figure*}

Now let us assume that $\mathbf{C}$ falls into two contributions $\mathbf{A}$ and $\mathbf{B}$ and that only $\mathbf{B}$ is mischaracterized by a constant factor in our model,
\begin{equation}
\mathbf{M} = \mathbf{A} + \alpha\mathbf{B}\ .
\end{equation}
Let us furthermore assume that $\mathbf{B}$ has a dominant eigenvalue $\lambda$ and that $\mathbf{v}$ is an eigenvector to it. If
\begin{equation}
\left| \lambda \mathbf{v} - \mathbf{C} \mathbf{v} \right| \ll \left| \lambda \mathbf{v}\right|
\end{equation}
then the matrix $\mathbf{C}$ and -- for values of $\alpha$ that are not too small -- also the matrix $\mathbf{M}$ will have an eigendimension close the that of $\mathbf{B}$ with eigenvalues $\lambda_C \approx \lambda$ and $\lambda_M \approx \alpha \lambda$. As a consequence, the matrix $\mathbf{X}$ will have an eigendimension with eigenvalue close to $\lambda_X \approx \frac{1-\alpha}{\alpha}$ which allows the same conclusion in \ref{sec:rescaling_of_cov}.

In section \ref{sec:examples} we considered a deformation of the halo model covariance of the form $\mathbf{M} = \mathbf{A} + \mathbf{B}_m$ with
\begin{equation}
\mathbf{B}_m = \alpha\mathbf{C}^{ss, \mathrm{Gauss}} + \beta\left( \mathbf{C}^{ss, \mathrm{halo}} - \mathbf{C}^{ss, \mathrm{Gauss}}\right)\ .
\end{equation}
This is similar to the situation described above. To illustrate how the rescaling factors $\alpha$ and $\beta$ impact the convergence of the PME we can e.g. compare the Fisher contours derived from $\mathbf{C}^{-1}$, $\mathbf{M}^{-1}$, $\mathbf{\Psi}^{1st}$ and $\mathbf{\Psi}^{2nd}$. In Fig. \ref{fi:cons_multi} we show the $1\sigma$ and $2\sigma$ Fisher-contours for the parameter pair $\Omega_m$-$\sigma_8$ derived for the DES multi-probe data vector using different values of $\alpha$ and $\beta$. The figure shows that the PME manages to correct the bias between contours derived from $\mathbf{C}^{-1}$ and contours derived from $\mathbf{M}^{-1}$ even for rather drastic choices of the rescaling factors. Especially for $\alpha,\ \beta > 1.0$ the convergence is very robust. As predicted by our considerations above, it however breaks down for $\alpha,\ \beta < 0.5$ where one can see strong oscillations between $\mathbf{\Psi}_{1\mathrm{st}}$ and $\mathbf{\Psi}_{2\mathrm{nd}}$. The convergence of the contours in Fig. \ref{fi:cons_multi} is very similar when other parameter combinations are considered or when the contours are derived for the other data vectors considered in this paper.

\subsubsection{Log-normal motivated approximation to the halo-model covariance}

Motivated by the work of \citet{Hilbert2011} on approximating the shear-shear covariance matrix with a log-normal approach (cf. their equation 26) we approximate the non-Gaussian parts of the covariance of shear correlation functions as
\begin{equation}
\label{eq:log-normal_scaling}
\langle \Delta \xi_{\pm}^A(\theta_i)\Delta\xi_{\pm}^B(\theta_j) \rangle_{\mathrm{non}\ \mathrm{Gauss.}} = \xi_{\pm}^A(\theta_i)\xi_{\pm}^B(\theta_j) R_{AB}
\end{equation}
where $\theta_i$ labels the different angular bins, $A$ and $B$ label the different auto- and cross-correlation functions and $R_{AB}$ is just a constant factor (depending only on the pair $A,B$ and not on whether $\xi_+$ or $\xi_-$ are involved). We fix the values of $R_{AB}$ by demanding that our approximation coincides with the halo-model for $\langle \Delta \xi_{+}^A(\theta)\Delta\xi_{+}^B(\theta) \rangle_{\mathrm{non}\ \mathrm{Gauss.}}$ where $\theta$ is a certain angular scale which we chose to be either our smallest angular bin ($\theta \approx 3'$) or a slightly larger scale ($\theta \approx 20'$). Note that this is a very crude approximation -- even to the log-normal model by \citet{Hilbert2011} since they have not even considered cross-correlations between redshift bins.

\begin{figure}
  \includegraphics[width=0.45\textwidth]{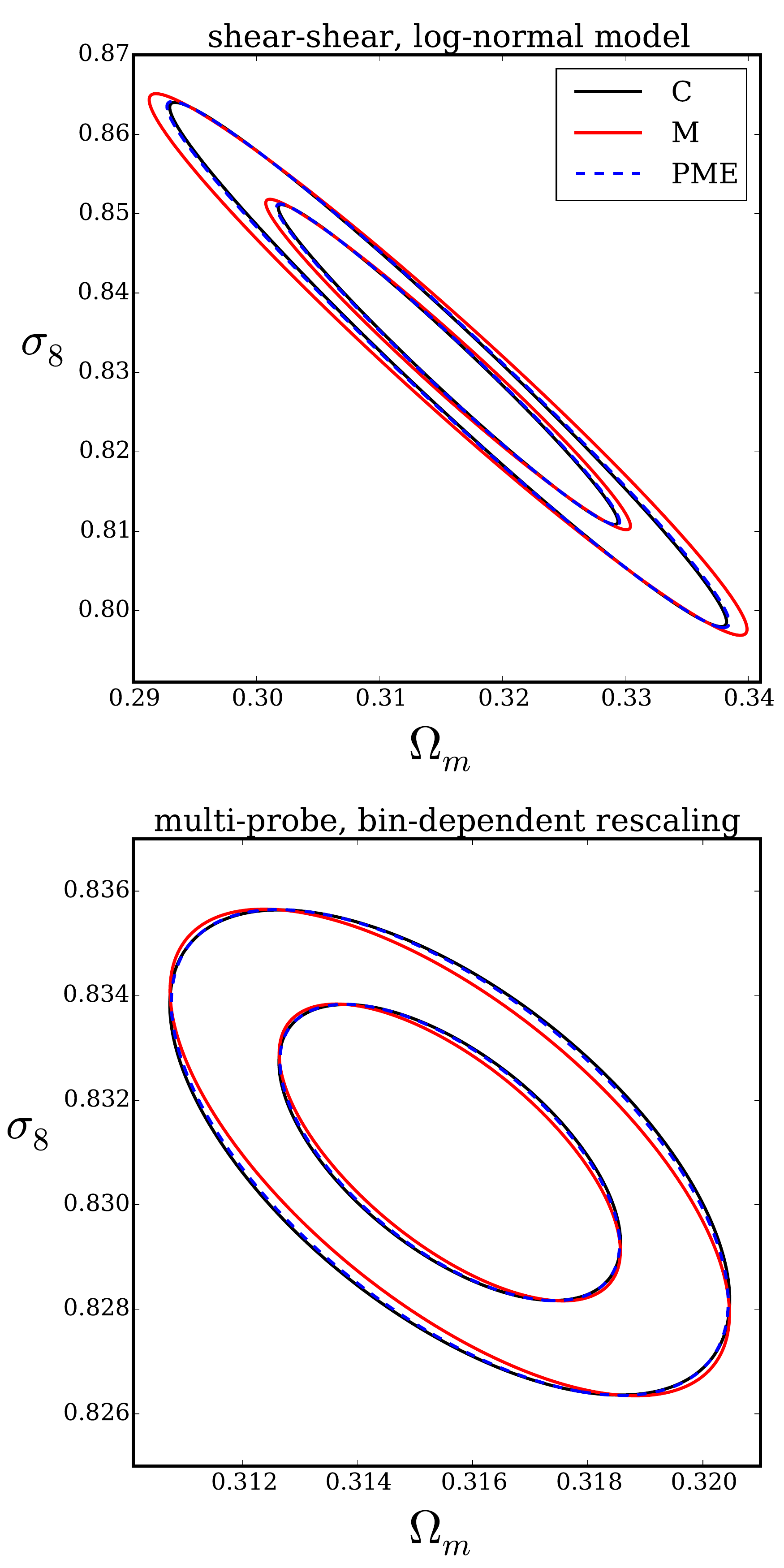}
   \caption{Top: $1\sigma$ and $2\sigma$ Fisher contours in the $\Omega_m$-$\sigma_8$ plane for the DES weak lensing data vector. The black contours are derived from our fiducial halo-model covariance $\mathbf{C}$. For the red contours we used a model covariance $\mathbf{M}$ that was motivated from the general structure of the log-normal covariance model for shear-shear correlation functions by \citeauthor{Hilbert2011} (2011, see main text). The PME (blue dashed contours) still manages to correct for the deviation between the two models. It should however be noted that in this case one eigenvalue of the deviation matrix $\mathbf{X}$ comes dangerously close to one ($\lambda_{\max} = 0.966$). As we discuss in the main text, this situation stabilizes if we match the amplitudes of the halo-model and the log-normal motivated covariance at intermediate angular scales ($\theta \sim 20'$) instead of the smallest scale of our data vector ($\theta \sim 3'$). Bottom: we applied a scale dependent rescaling of the halo-model covariance for the multi-probe data vector motivated by findings of \citet{Friedrich2015}. The PME converges also in this case.}
  \label{fi:appendix_1}
\end{figure}

We nevertheless use the above matrix as our model covariance $\mathbf{M}$ for the DES shear-shear data vector and compare it to the halo-model covariance $\mathbf{C}$ and the PME. All eigenvalues of matrix $\mathbf{X}$ have in that case $|\lambda_i| < 1$. The three most dominant eigenvalues are
\begin{eqnarray}
\lambda_1 &=& 0.776 \nonumber \\
\lambda_2 &=& -0.675 \nonumber \\
\lambda_3 &=& 0.197 \nonumber \\
\end{eqnarray}
in the case where we match the amplitudes of $\mathbf{M}$ and $\mathbf{C}$ at $\theta \approx 20'$ and
\begin{eqnarray}
\lambda_1 &=& 0.966 \nonumber \\
\lambda_2 &=& -0.442 \nonumber \\
\lambda_3 &=& 0.203 \nonumber \\
\end{eqnarray}
when we match the amplitudes at $\theta \approx 3'$. In both of these cases the PME in principle converges. However, in the second case at least one eigenvalue comes dangerously close to $1$. In Fig. \ref{fi:appendix_1} we show that in terms of the Fisher contours in the $\Omega_m$-$\sigma_8$-plane the PME  nevertheless converges and significantly corrects for the deviations between halo-model and log-normal motivated covariance. We have also checked other parameter combinations and find similar results. The reason that a matching at larger scales gives smaller eigenvalues (i.e. better agreement between halo-model and log-normal motivated covariance) is probably that the scaling of Eq. \ref{eq:log-normal_scaling} fails at small scales.

\newpage

\subsubsection{Scale dependent rescaling of the cosmic variance of the multi-probe data vector}

Another alternative way to deform the halo-model covariance is to apply different rescaling factors $\alpha$ and $\beta$ for the Gaussian and non-Gaussian cosmic variance parts for different angular scales (cf. Eq. \ref{eq:definition_of_pretend_model}). If e.g. the finite area of a survey is not correctly accounted for in a covariance model, the results of \citet{Friedrich2015} indicate that this leads to a scale dependent miss-characterization of the Gaussian cosmic variance and to an almost scale independent over- or underestimation of the non-Gaussian parts. Covariance parts involving shape- or shot-noise on the other hand are less sensitive to the survey area (only to the product of area and galaxy density which is the total number of galaxies).

Motivated by this we replace Eq. \ref{eq:definition_of_pretend_model} by
\begin{equation}
\mathrm{B}_{m, ij} = \alpha_{ij}\mathrm{C}_{ij}^{ss, \mathrm{Gauss}} + \beta\left( \mathrm{C}_{ij}^{ss, \mathrm{halo}} - \mathrm{C}_{ij}^{ss, \mathrm{Gauss}}\right)
\end{equation}  
where we choose $\beta = 0.5$ and $\alpha_{ij} = \sqrt{a_i a_j}$ setting $a_i$ to $1.0$ at the smallest scales and to $0.5$ at the largest scales of the data vector and linearly interpolating for intermediate bins (interpolating in terms of the bin-index).

The most dominant eigenvalues of the matrix $\mathbf{X}$ for this choice of the matrix $\mathbf{M}$ are
\begin{eqnarray}
\lambda_1 &=& 0.709 \nonumber \\
\lambda_2 &=& -0.440 \nonumber \\
\lambda_3 &=& 0.242\ , \nonumber \\
\end{eqnarray}
i.e. the PME converges. The bottom panel of Fig. \ref{fi:appendix_1} also shows that the Fisher contours derived from the 2nd order PME around this model almost coincide with the ones derived from the halo-model covariance again.

\subsection{Convergence in the General Case}

Let us now consider the general case. We want invert the equation
\begin{equation}
\label{eq:appendix_1}
\mathbf{C} = \mathbf{M}^{1/2}\left(\mathbb{1} + \mathbf{X}\right)\mathbf{M}^{1/2}
\end{equation}
where
\begin{equation}
\mathbf{X} := \mathbf{M}^{-1/2}(\mathbf{B}-\mathbf{B}_m)\mathbf{M}^{-1/2}\ .
\end{equation}
Since both $\mathbf{M}$ and $\mathbf{C}$ are positive definite matrices we can immediately infer that also the matrix $\mathbb{1} + \mathbf{X}$ must be positive definite, i.e. all its eigenvalues must be greater that $0$. As a consequence, all eigenvalues $\lambda_i$ of $\mathbf{X}$ must fulfil
\begin{equation}
\lambda_i > -1\ \forall i\ .
\end{equation}
In order to invert $\mathbb{1} + \mathbf{X}$ let us change into the eigenbasis of $\mathbf{X}$ by means of an orthogonal matrix $\mathbf{U}$, i.e.
\begin{equation}
\mathbb{1} + \mathbf{X} = \mathbf{U}^T \mathrm{diag}(1 + \lambda_i) \mathbf{U}\ .
\end{equation}
It is not a priori clear whether we can invert this by means of the geometric series, since we do not know a priori that $|\lambda_i| < 1$. As discussed in section \ref{sec:conclusions} in the case that $|\lambda_i| > 1$ the PME can at least help to identify differences between a covariance model and covariance from (possibly very few) simulations since in that case the 1st order and 2nd order PME will display a divergent behaviour. However, since we know a priori that $\lambda_i > -1$ we can in principle apply a trick to let the PME series converge in any case. This trick is to expand $1/(1+\lambda)$ not around $\lambda_0 = 0$ but around some other point $\lambda_0 = a > 0$:
\begin{equation}
\frac{1}{1+\lambda} = \frac{1}{1+a} \left[1 - \left(\frac{x-a}{1+a}\right) + \left(\frac{x-a}{1+a}\right)^2 -\ \cdots \right]\ .
\end{equation}
In terms of the PME series this is in fact equivalent to replacing the model covariance $\mathbf{M}$ by $(1+a)\mathbf{M}$. This way, one can in principle always ensure convergence of the series. This however comes at the expense of the series converging very slowly for eigenvalues of $\mathbf{X}$ that are already close to or smaller than $0$. Since in a real case scenario $\mathbf{M}$ is assumed to be our best guess for the true covariance we hence recommend to stay with $a=0$ and interpret a divergent PME as a significant difference between model and N-body covariance.

\section{Data vectors}
\label{app:data_vector}

\subsection{Weak lensing data vectors}

The redshift distribution and tomographic binning used for our LSST-like weak lensing data vector was chosen to be exactly that of \citet[see section 3]{Krause2016}. This means we assumed an overall source density of $26\mathrm{arcmin}^{-2}$ and a source distribution with a median redshift of $\approx 0.7$ that extends out to $z \gtrsim 3.0$. The tomographic bins were defined by first splitting the redshift distribution into $10$ non-overlapping bins of equal source density and then assuming a Gaussian photoz uncertainty of $\sigma_z = 0.05$. The intrinsic ellipticity dispersion of the sources was assumed to be $\sigma_\epsilon = 0.26$ per ellipticity component.

The redshift distribution for the DES-like data vector was chosen to be shallower as for the LSST case reflecting the smaller depth of DES. Here our source distribution has a median redshift of $\approx 0.5$ and extends out to $z = 2.0$. The overall source density was taken to be $10/\mathrm{arcmin}^2$ and the $5$ tomographic bins where defined assuming a photoz uncertaintz of $\sigma_z = 0.08$. The intrinsic ellipticity dispersion was chosen to be the same as for the LSST-like case.

\subsection{Lens galaxies}

For the DES multi-probe data vector we also considered galaxy clustering and galaxy-galaxy lensing correlation functions. For this we were assuming a sample of foreground galaxies with a constant comoving density motivated by the DES redMaGiC sample \citep{Rozo2016} divided into $3$ tomographic bins whose redshift ranges are $(0.20, 0.35)$, $(0.35, 0.50)$ and $(0.50, 0.65)$. For these galaxies we assumed zero redshift uncertainties motivated by the fact that the redMaGiC redshift errors are small compared to the values for our source samples. The overall density of forground galaxies was taken to be $0.15/\mathrm{arcmin}^2$.

\subsection{Binning and scales}

The real space data vectors use 15 logarithmic angular bin from $\theta = 2.5'$ to $\theta = 250'$ for each correlation function and the Fourier space data vector uses 40 logarithmic bin from $\ell = 20$ to $\ell = 5000$ for each power spectrum. Data vector I contains the correlation functions $\xi_+$ and $\xi_-$ for each possible combination of source bins. Data vector Ia also contains the auto-correlation of the lens bins and all possible combinations of lens-source correlations (i.e. only those combinations where the sources are at higher redshifts than the lenses). Data vector II contains the auto- and cross-power spectra of all possible combinations of source bins.

\end{document}